\documentclass[12pt]{article}

\usepackage{amsmath,amsfonts,amssymb}
\newcommand{\be}{\begin{equation}}
\newcommand{\ee}{\end{equation}}
\newcommand{\bea}{\begin{eqnarray}}
\newcommand{\eea}{\end{eqnarray}}
\newcommand{\ba}{\begin{array}}
\newcommand{\ea}{\end{array}}
\newcommand{\bse}{\begin{subequations}}
\newcommand{\ese}{\end{subequations}}

\makeatletter \@addtoreset{equation}{section}

\makeatother
\begin{document}

\baselineskip 18pt%

\begin{titlepage}
\vspace*{1mm}%
\hfill%
\vbox{
    \halign{#\hfil \cr
\;\;\;\;\;\;\;\;\;\; IPM/P-2008/015 \cr
           \;\;\;\;\;\;\;\;\;\;\;\;\;\; SUT-P-08-2a   \cr
   \;\;\;\;\;\;\;\;\;\;\;\;\;\; IC-2008-017 \cr
arXiv:0805.0203 {\tt [hep-th]} \cr
           } 
      }  

\centerline{{\Large {\bf  \textsl{Nearing  $11d$ Extremal Intersecting Giants}}}}%
\centerline{{\large {\bf and}}}
\centerline{{\Large {\bf \textsl{New Decoupled Sectors in $D=3,\ 6$ SCFTs  }}}}%
\vspace*{5mm}
\centerline{{\bf \large{R. Fareghbal$^{1,2}$, C. N. Gowdigere$^{3}$,
A.E. Mosaffa$^{1}$, M. M. Sheikh-Jabbari$^{1}$}}}
\begin{center}
{\it {$^1$School of Physics, Institute for Research in Fundamental Sciences  (IPM)\\
P.O.Box 19395-5531, Tehran, IRAN\\
$^2$Department of Physics, Sharif University of Technology\\
P.O.Box 11365-9161, Tehran, IRAN\\
$^3$ The Abdus Salam ICTP, Strada Costiera 11, Trieste Italy }}

{E-mails: {\tt fareghbal, mosaffa, jabbari @theory.ipm.ac.ir, cgowdige@ictp.it}}%
\end{center}

\begin{center}{\bf Abstract}\end{center}
\begin{quote}

 We extend the  analysis of arXiv:0801.4457 [hep-th] to charged black hole solutions of four-dimensional $U(1)^4$  gauged
supergravity which carry three charges. There are two decoupling
near-horizon limits, one over the near-BPS black hole solution and
the other over the near-extremal, but non-BPS geometry. Taking the
limit over the eleven dimensional uplift of these black hole
solutions,  for both of these cases we obtain a geometry which has
a piece (conformal) to rotating BTZ$\times S^2$. We study the
$4d$, $11d$ and $5d, 3d$ black hole properties. Moreover, we show
that the BTZ$\times S^2$ geometry obtained after the near-BPS
(near-extremal) limit is also a solution to five-dimensional
$U(1)^3$ un-gauged (gauged) STU supergravity. Based on these
decoupling limits we argue that there should be sectors of $3d$
CFT resulting from low energy limit of theory on $N$ M2-branes
($N\to \infty)$, which are decoupled from the rest of the theory
and are effectively described by a $2d$ CFT. The central charge of
the $2d$ CFT in both near-BPS and near-extremal case scales as
$N$. The engineering dimension of the operators in these decoupled
sectors scales as $N^{4/3}$ (for near-BPS case) while as $N^{3/2}$
(for the near-extremal case). Moreover, we discuss the description
of the decoupled sectors as certain deformations of $6d$ CFT
residing on the intersecting M5-brane giants.

\end{quote}
\end{titlepage}
%
%
\tableofcontents

\section{Introduction and Summary}\label{Introduction-section}

Understanding the statistical mechanical origin of black hole
thermodynamics has been an ever challenging question posed to string
theory. In this regard BPS or extremal black holes  in AdS
background in various dimensions have been of particular interest,
as via AdS/CFT \cite{MAGOO} we have the possibility of a
description for black hole microstates through the dual CFT operators. Here we focus on a specific class of black hole solutions to ${\cal N}=2$, $d=4$ $U(1)^4$ gauged
supergravity \cite{Cvetic:1999xp,Duff:1999rk}. These solutions can
be uplifted to $11d$ as (black brane) deformations to $AdS_4\times
S^7$ geometry by the addition of  various intersecting spherical
M5-brane giant gravitons \cite{11d-superstars, giant} and their
excitations. These M5-branes wrap various five-spheres inside the
$S^7$ \cite{11d-superstars}.

Since the $11d$ description of these $4d$ charged black holes is as
deformations of $AdS_4\times S^7$, these black holes should also have a
description in the $3d$ dual CFT. Although our knowledge of this
CFT is very limited, we may still study certain sectors of this
theory. For example the BMN sector of this $3d$ CFT is  described
by the plane-wave matrix model (also known as BMN matrix model)
\cite{BMN, DSV}, which in turn is the DLCQ of M-theory on $AdS_4 \times S^7$ \cite{DLCQ-Penrose, TGMT}.  In this work, among other things, we make the
first steps in gaining a better understanding of certain sectors of
the dual $3d$ CFT by relating it to better manageable theories
like $2d$ CFT's.

Here we focus on the three-charge black hole solutions to $U(1)^4$
$4d$ gauged supergravity and argue that the near-extremal black hole
solutions admit near-horizon  decoupling limits to a geometry which
has $X_{M,J}\times S^2$,  $X_{M, J}$ being global $AdS_3$, or
$AdS_3$ with conical singularity or a rotating BTZ black hole.
Through out the paper we will discuss the limit from the $4d$, $11d$
and $5d$ gravity theories as well as the $3d$ and $2d$ CFT's (noting
the appearance of $AdS_3$ factor in the decoupled geometry). In this
way we support our statement about existence of decoupled sectors.
This is an M-theory version of the ``two-charge'' black holes
discussed in \cite{FGMS-1}, see also \cite{Balasubramanian:2007bs}.
Similar to the ``two-charge'' black hole of \cite{FGMS-1}, there are
two decoupling limits. The first is a near-horizon limit on the
near-extremal black hole while also going to the near-BPS limit; we
will refer to this near-horizon near-extremal limit as simply the
``near-BPS'' limit. The second is a near-horizon limit on the
near-extremal black hole and continuing to be far-from-BPS; we will
refer to this near-horizon near-extremal limit as the
``far-from-BPS'' limit. Throughout this work, we discuss the
near-BPS and far-from-BPS  cases in parallel.

This paper is organized as follows. In section \ref{Section2}, we
first review generic charged black hole solutions to $4d$ $U(1)^4$
gauged SUGRA, and discuss their uplift to $11d$ supergravity where
they appear as black five-brane deformations of $AdS_4\times S^7$.
In section \ref{Section3-Limits}, we focus on the three-charge black
holes when they are close to saturating the extremality  bound. We
perform both the near-horizon limits i.e. the near-BPS and the
far-from-BPS ones. In both cases we obtain a  $11d$ geometry
containing an $AdS_3\times S^2$ (or more generally  a static
BTZ$\times S^2$) factor. Then we turn on the fourth charge in a
perturbative manner, that is we choose the fourth charge to be much
smaller than the other three. The effect of the perturbative
addition of the fourth charge in the near-horizon decoupling limits
is the addition of angular momentum to the static BTZ black hole, to
obtain a rotating BTZ black hole.

In section \ref{Section4-5d-gravity}, we show that the BTZ $\times
S^2$ geometries obtained in the near-horizon limits are indeed
solutions to certain $5d$ supergravities; that of the near-BPS case
is a solution to ungauged $5d$ STU model \cite{Gunaydin:1983bi}
while that of the far-from-BPS geometry is a solution to the $5d$
gauged $U(1)^3$ supergravity \cite{Gunaydin:1984ak}.

In section \ref{Section5-entropy}, we analyze the Bekenstein-Hawking
entropy of the $4d$ black holes, the near-horizon limits of which
was discussed in section 3. We show that for both the near-BPS and
far-from-BPS cases the $4d$ entropy and the entropy of the rotating
BTZ black hole obtained after the limit are exactly matching.

In  section \ref{Section-dual-CFTs}, we present dual CFT pictures of
the gravity analysis of the previous sections. The near-horizon
limit in the language of $3d$ CFT (dual to M-theory on $AdS_4$ with
$N$ units of four-form flux) translates into $N\to \infty$ limit,
while focusing on specific BMN-type sectors of the $3d$ CFT. In the
near-BPS case this sector is identified with operators which carry
three R-charges $J_i$ of the dual CFT with $\Delta\sim J_i\sim
N^{4/3}\to \infty$ ($\Delta$ is the engineering dimension of the
operators), while $\Delta-\sum_i J_i\sim N$. In the far-from-BPS
limit, however, we are dealing with operators with $\Delta\sim
J_i\sim N^{3/2}$, while a certain combination of $\Delta,\ J_i$ (see
\eqref{Ext-bound}) scales as $N$. Due to the presence of $AdS_3$
factors in the decoupled geometry, one then expects the BMN-type
operators to have a description in terms of $2d$ CFT's. We briefly
discuss the corresponding $2d$ CFT's, and read the central charge of
the theory, which in both the near-BPS and far-from-BPS cases scales
as $N$. In addition, we also give a mapping between the $3d$ CFT
charges, $\Delta, J_i$ and the $L_0$ and $\bar L_0$ of the $2d$
CFT's.

The last section is devoted to discussions and concluding remarks.
In appendix \ref{appendixA}, we have gathered computations showing that our
decoupled geometries after the limit are indeed solutions to $11d$
supergravity. In appendix \ref{appendixB}, we show that running the entropy function machinery \cite{Sen-review} for the BTZ $\times S^2$ geometries correctly reproduces the black hole entropy.

\section{Charged Black Holes in $4d$ $U(1)^4$ SUGRA}\label{Section2}

The black hole solutions that we are interested in and will be
reviewed in this section are the static charged solutions to ${\cal
N} = 2$ $U(1)^4$ gauged supergravity in four dimensions which were
first obtained in \cite{Sabra:1999ux, Duff:1999gh} (see
\cite{Duff:1999rk} for a review). These solutions can be uplifted to
eleven dimensions \cite{Cvetic:1999xp}. They were analyzed in
\cite{11d-superstars} where it was discussed that they correspond to
condensates of intersecting smeared (delocalized) M-theory spherical
M5-brane giant gravitons and referred to as superstars/black holes.
As solutions to $11d$ supergravity they are specified by their
metric and the three-form field:
\begin{equation}\label{11-metric-original}
ds_{11}^2= \Delta^{\frac23}  \left[-\frac{f}{H}dt^2+\frac{dr^2}{f} +r^2\,d \Omega_2^2\right]\\
+\Delta^{-\frac13} \left[\sum_{i=1}^4L^2
H_i\left(d\mu_i^2+\mu_i^2\left[d\phi_i+a_i\
dt/L\right]^2\right)\right],
\end{equation}
\begin{equation}\label{11d-3form}
C^{(3)} = -\frac{r^3}{2} \Delta\, dt \wedge d^2\Omega_2 -
\frac{L^2}{2} \sum_{i=1}^4 {\tilde q}_i \,\mu_i^2 \,\left( d\phi_i -
\frac{q_i}{{\tilde q}_i} \frac{dt}{L} \right) \wedge d^2\Omega_2.
\end{equation}
In the above
\be\label{11d-metric-Handf}\begin{split}%
H=H_1H_2H_3H_4, \qquad H_i &= 1 +  \frac{q_i}{r} ,\qquad f = 1 - \frac{\mu}{r} + \frac{4\,r^2}{L^2} H,   \\
\Delta = H
\bigl[\frac{\mu_1^2}{H_1}+\frac{\mu_2^2}{H_2}+\frac{\mu_3^2}{H_3} &
+\frac{\mu_4^2}{H_4}\bigr],
\qquad a_i = \frac{{\tilde q}_i}{q_i}  \left[ \frac{1}{H_i}-1 \right], \\
\mu_1 = \cos \theta_1, \quad \mu_2 =  \sin \theta_1 \cos \theta_2 ,
\quad \mu_3 =& \sin \theta_1 \sin \theta_2 \cos \theta_3, \quad
\mu_4 = \sin \theta_1 \sin \theta_2 \sin \theta_3,
\end{split}\ee%
and $d\Omega_2^2$ and $d^2\Omega_2$ are respectively the metric and
the volume form on a unit radius two-sphere.

These geometries asymptote to, i.e. as  $r\rightarrow \infty$,
$AdS_4 \times S^7$ with radii $L/2$ and  $L$ respectively, and
$\theta_1, \theta_2, \theta_3,  \phi_1, \phi_2, \phi_3, \phi_4$
parameterize the angles of the seven-sphere. They constitute a five
parameter family of solutions, specified with $\mu, q_i,\ i=1,2,3,4$
and
\begin{equation}\label{tilde-qi}
{\tilde q}_i = \sqrt{q_i (  q_i+ \mu)}\ .
\end{equation}
From the metric and the three-from expressions it is evident that in
our conventions the parameters, $\mu,\ q_i,\ \tilde q_i$ and $L$,
are all of dimension of length.

Upon a specific reduction of the $11d$ supergravity on $S^7$
\cite{Cvetic:1999xp}, the above geometries are mapped onto the
electrically charged black hole solutions of $4d$ ${\cal N}=2$
$U(1)^4$ gauged supergravity \cite{Sabra:1999ux} which are also
solutions of ${\cal N}=8$ $SO(8)$ gauged supergravity \cite{Duff:1999gh}:%
\begin{subequations}\label{4dBH}
\begin{align}
ds_{4}^2&=-H^{-1/2} f\ dt^2+ H^{1/2}\left(\frac{dr^2}{f}+r^2d\Omega_{2}^2\right),  \\
 A_i &=\frac{\tilde q_i}{q_i} \left(\frac{1}{H_i}-1\right)dt, \quad X_i=\frac{H^{1/4}}{H_i},
\end{align}
\end{subequations}
where $f$, $H_i$ and $H$ are given in \eqref{11d-metric-Handf} and
$A_i$ and $X_i$ are parameterizing the four gauge fields and three
scalars of the four-dimensional theory (note that $X_1X_2X_3X_4=1$).
The physical observable charges associated with these $4d$ black
holes are the ADM mass \cite{Sabra-Liu}
\begin{equation}\label{4d-ADM-mass}
M = \frac{1}{2 G_N^{(4)}} (2 \mu + q_1 + q_2 + q_3 + q_4 ).
\end{equation}
and electric charges%
\be\label{angular-momenta}
 J_i=\frac{L}{2  G_N^{(4)}} \tilde q_i\ .
\ee%
In the above expressions $G_N^{(4)}$ is the four-dimensional Newton
constant which is related to the $11d$ Newton constant as
\begin{equation}\label{GN-4-11}
G_N^{(4)} =\frac{G^{(11)}_N}{\frac{\pi^4}{3}{L^7}}=
\frac{3}{8\sqrt2}\frac{L^2}{N^{3/2}}, \qquad G_N^{(11)}
=16\pi^7l_p^9,\qquad L^6=32\pi^2 l_p^6 N\ .
\end{equation}
As argued from the $11d$ viewpoint these geometries correspond to
(at most) four stacks of intersecting spherical five-branes; each
stack carries  angular momentum $J_i$ \cite{Cvetic:1999xp} and
consists of $N_i$ number of M5-branes \cite{11d-superstars}%
\be\label{five-brane-number}%
N_i=\frac{3 J_i}{4N}=\sqrt2 N^{1/2} \frac{\tilde q_i}{L}. %
\ee%
The above number has been computed noting that each M5-brane carries
one unit of the flux of the four-form field strength $F_4=dC_3$, explicitly%
\be\label{Ni-M5-flux}%
 N_i=\frac{1}{16\pi G^{(11)}_N T_{M_5}}\ \int
F_4,\qquad T_{M_5}=\frac{1}{(2\pi)^5 l_p^6}, %
\ee%
 and noting the expression for the three-form \eqref{11d-3form}.

For generic $\mu> 0$ the above solutions are non-BPS breaking all
the supersymmetries, for $\mu=0$, however, the solution is  BPS. For
a solution with $n$ number  of non-zero charges $q_i$, the BPS
solutions, as solutions to $11d$ supergravity, preserve $32/2^n$
number of supersymmetries. In the four-dimensional setting black
holes with regular horizons can occur only when $\mu \neq 0$ and
hence are all non-supersymmetric. The supersymmetric solution
corresponds to a naked singularity and hence were termed superstars
in \cite{11d-superstars}. Although in this paper we will mainly be
interested in the three-charge and four-charge M-theory
superstar/black holes, for completeness we  present a brief account
of each of one, two, three and four charge black holes separately.

\begin{itemize}
\item{\textbf{One-charge black hole:}} For $\mu=0$ we have a space
with naked, null  singularity which preserves 16 supercharges. The
eleven-dimensional LLM solutions \cite{LLM} are  de-singularized
counter-parts of these solutions. As soon as we turn on $\mu$ we
obtain a space-like singularity at $r=0$ which is hidden behind a
regular horizon.

From the $11d$ viewpoint these solutions correspond to M5-branes
wrapping $S^5$ inside the $S^7$, while rotating on a circle on
$S^7$.  The spherical M5-branes, the giant five-branes,  are smeared
on the two directions transverse to their worldvolume. The
non-extremality parameter $\mu$ corresponds to turning on (membrane)
excitations on  the five-brane. These excitations are such that they
do not change the $S^5$ shape of the five-brane.

\item{\textbf{Two-charge black hole:}} For $\mu=0$ we again have a
null, naked singularity and the background is 1/4 BPS. For any
$\mu>0$ we have a space-like singularity at $r=0$ and a horizon
sitting at the larger root of function $f$.

In the $11d$ picture we have two stacks of spherical five-branes
which are intersecting on an $S^3$ inside $S^7$, while each moving
on a circle and are smeared on directions transverse to their
worldvolume.

\item{\textbf{Three-charge M-theory superstar/black hole:}
When only one of the four charges, say the $q_1$, is vanishing, the
$4d$ black hole  has a different causal and singularity structure
than the one and two charge cases.
 For $0\leq\mu < \mu_c$, with%
\be\label{muc-3charge}%
\mu_c=\frac{4}{L^2} q_2q_3q_4,%
\ee%
we have a solution with a time-like singularity sitting at $r=0$ and
no regular horizon. This case resembles a standard
Reissner-Nordstrom black hole which violates the extremality bound.
For the $\mu=0$ case the solution becomes BPS, preserving 4
supercharges. For $\mu=\mu_c$ the solution is ``extremal'', but
non-supersymmetric (non-BPS),  with a vanishing horizon size.
Similar to the extremal Reissner-Nordstrom solution, the geometry
has a null singularity. When $\mu>\mu_c$ we have a regular
Reissner-Nordstrom-type black hole with a finite size horizon and a
space-like singularity sitting behind the horizon.

In the eleven-dimensional viewpoint we have three stacks of smeared
spherical M5-branes on $S^7$ while interesting on an $S^1$. As we
will argue one may interpret the stack of intersecting giants at
$\mu=\mu_c$ as a (non-marginal) bound state of M5-branes which are
wrapping (holomorphic) 4-cycles $\Sigma_4$ on a $CP^3$, and hence
with worldvolume $R\times S^1\times \Sigma_4$. Turning on the
non-extremality parameter $\mu>\mu_c$ is then like turning on the
five-brane type excitations on the system of intersecting giants.}

\item{ \textbf{Four-charge M-theory superstar/black hole:}
As far as the causal and singularity structures are concerned, this
case is very similar to the three charge case, with two differences.
First, $\mu_c$ has now a different complicated expression in terms
of $q_i$'s than \eqref{muc-3charge}, but in any case $\mu_c\geq
\frac{4}{L^2}(q_1q_2q_3+q_1q_2q_4+q_1q_3q_4+q_2q_3q_4)$. Second, in
this case at $\mu=\mu_c$, unlike the three-charge case, we have an
extremal solution with non-vanishing horizon size. Note that in this
case one can extend the geometry past $r=0$ to $r\geq -q_1$, where
$q_1$ is the smallest of the four charges. For $ 0\leq \mu<\mu_c$
the time-like naked singularity is sitting behind $r=0$, at
$r=-q_1$.

In terms of M5-branes, the $11d$ geometry corresponds to four stacks
of smeared M5-brane giants their worldvolume intersecting only on
the time direction.}
\end{itemize}

\section{Near-Horizon Limits of Three-Charge Black
Holes}\label{Section3-Limits}

Among the black hole solutions reviewed in the previous section, the
three-charge case has certain unique features.  In particular, for a
given set of three non-zero charges, there are two possible
near-extremal limits. The first is what we have already referred to
as near-BPS case; the near-extremal black holes thus obtained have
$\mu \sim 0$ with  $\mu_c/L\ll 1$. The second is what we have
already referred to as the far-from-BPS limit. The near-extremal
black holes thus obtained have  $\mu \sim \mu_c$ \eqref{muc-3charge}
with with $\mu_c\sim L$. Both these limits result in vanishing
horizon area. The M-theory three-charge case that we consider here
is analogous to the IIB two-charge case discussed in \cite{FGMS-1}.

In taking the near-horizon limit of the
three-charge extremal black hole in four dimensions, we face the
same problems as we did for the five-dimensional two-charge extremal
black hole \cite{FGMS-1},  namely we do not obtain a product
geometry with $AdS$ and sphere factors. However, working with the
uplifted eleven dimensional solution, we do obtain a product geometry: $AdS_3
\times S^2$.

\subsection{The near-horizon near-BPS limit}

We require $\mu_c\rightarrow 0$ together with
$r\rightarrow 0$. More precisely, we consider $\epsilon\to 0$ with
the following scalings

 $\bullet$ \emph{$\mu_1\sim 1$ case}
\begin{equation}\label{near-limit-BPS-1}
\begin{split}
 q_i=\epsilon \hat{q}_i\ \Rightarrow\ &\mu_c=\epsilon^3\hat{\mu_c}, \quad \hat\mu_c\equiv
 \frac{4\hat q_2\hat q_3 \hat q_4}{L^2},
\qquad \mu-\mu_c=\epsilon^3 M^2,\\
r=\epsilon^3 \hat\mu_c \rho^2,&\qquad \mu_i= \epsilon^{1/2}
x_i,\quad  \psi_i=\phi_i-t/L,\  (i=2,3,4)
\end{split}
\end{equation}
while keeping $\rho,\ \hat\mu_c,\ \hat q_i,\ M,\ x_i, \psi_i, \ L$
fixed. Note also that, as $\mu_1^2=1-\mu_2^2-\mu_3^2-\mu_4^2$, in
this limit $\mu_1=1+{\cal O}(\epsilon^2)$. This limit corresponds to
$\theta_1\sim \epsilon^{1/2}, \theta_2,\theta_3=$fixed \emph{cf.}
\eqref{11d-metric-Handf}.\\

$\bullet$ \emph{$\mu_1\sim \mu_1^0\neq 1$ case}
\be\label{near-BPS-limit-2}
\begin{split}
q_i=\epsilon \hat{q}_i\ \Rightarrow\ \mu_c=\epsilon^3\hat{\mu_c},
\quad \hat\mu_c\equiv \frac{4\hat q_2\hat q_3\hat q_4}{L^2},&
\qquad \mu-\mu_c=\epsilon^3 M^2,\\
r=\epsilon^3 \hat\mu_c \rho^2,  \qquad
\theta_i=\theta_i^0-\epsilon\hat{\theta}_i\ \Rightarrow\
d\mu_i=\epsilon \ d\hat{\mu}_i\ &, \qquad
 \psi_i=\frac{1}{\epsilon}(\phi_i-t/L),\ (i=2,3,4),
\end{split}
\ee%
where $0\le\theta_i^0\le\pi/2$ are fixed values for $\theta_i$
and $\rho,\ \hat{\theta}_i,\ \psi_i,\ \hat q_i,\ \hat\mu_c$ are kept
fixed.

Note that in both of the above limits the physical charges
$\tilde{q_i}\sim q_i\sim\epsilon$ and the function $f$ of
\eqref{11-metric-original} becomes
\begin{equation}\label{gamma}
f=1-\frac{\gamma^2}{\rho^2}.
\end{equation}
where
\begin{equation}
\gamma^2\equiv\frac{\hat{\mu}-\hat{\mu}_c}{\hat{\mu}_c}=\frac{\mu-\mu_c}{\mu_c}\
.
\end{equation}

Performing this limit on \eqref{11-metric-original}, we end up
with the following BTZ$\times S^2\times T^6$ metric
\begin{equation}\label{nearBPSmetric}
\frac{ds^2}{\epsilon^2}= (R^2_A\ ds_{BTZ}^2+ R^2_S
d\Omega_2^2)+\frac{L^2}{R_S}\ ds^2_{{\cal C}_6}
\end{equation}
where%
\begin{equation}\label{nearBPSAdS}
ds_{BTZ}^2=-(\rho^2-\gamma^2)\
d\tau^2+\frac{d\rho^2}{\rho^2-\gamma^2}+\rho^2d\phi_1^2
 \end{equation}
with $\tau=t/L$ and%
\be\label{Rs-Ra-BPS}%
R_A= 2R_S\ .%
\ee%
The radius of the two-sphere $R_S$ and the six-dimensional part
$ds^2_{{\cal C}_6}$ have different expressions for the two cases:

$\bullet$ \emph{$\mu_1\sim 1$ case}%
\be\label{Rs-mu=1}%
R^3_S=\hat q_2\hat q_3\hat q_4\ , \qquad ds^2_{{\cal
C}_6}=\sum_{i=2,3,4} \hat q_i (dx_i^2+x_i^2 d\psi_i^2)\ . %
\ee%

$\bullet$ \emph{$\mu_1\sim \mu_1^0\neq 1$ case}

\begin{equation}\label{Rs-mu-neq1}
R^3_S=(\mu_1^0)^2 \hat q_2\hat q_3\hat q_4 , \qquad ds^2_{{\cal
C}_6}=\sum_{i=2,3,4}\hat{q}_i\left(d\hat{\mu}_i^2+(\mu_i^0)^2d\psi_i^2\right).
\end{equation}

Upon appropriate periodic identifications, the ${\cal M}_6$ part
of the metric, in both of the above cases describes a $T^6$. For
$\gamma^2=-1$, the three dimensional part \eqref{nearBPSAdS}
describes a global $AdS_3$ space. Note that this corresponds to
$\mu=0$ i.e. the BPS point. For $-1<\gamma^2<0$, the space becomes
conical whereas for $\gamma=0$ we have a massless BTZ black hole.
Finally, for $\gamma^2>0$ the space is a massive BTZ black hole
with mass $\gamma R_A$. (For a concise review of our terminology
see Appendix A of \cite{FGMS-1}.)

Starting from the geometry caused by  three stacks of flat M5-branes,
intersecting on a $R^4$ and taking the near-horizon limit, results in a  BTZ$\times S^2\times T^6$ geometry  \cite{Vijay-Larsen}, which is essentially the geometry we obtained here. We will return to
this point later in section 6.

\subsection{The near-horizon far-from-BPS limit}

When  $\mu_c$ has a finite value, we are far from the BPS point. A
near-extremal black hole is described by $\mu-\mu_c\ll \mu_c\sim
L$. To take the near-horizon limit of such a solution we consider
the following scalings%
\be\label{near-ext-limit}%
\begin{split}%
&r=\frac{\epsilon^2\mu_c}{f_0}\ \rho^2, \qquad t=\frac{1}{\epsilon\sqrt{f_0}}\ \tau, \qquad  \phi_1=\frac{\varphi}{\epsilon},\\
&\mu-\mu_c=\epsilon^2\mu_cM, \qquad d\psi_i=
d\phi_i-\frac{\tilde{q_i}}{q_i}\frac{\tau}{\epsilon\sqrt{f_0}}\
(i=2,3,4),
\end{split}%
\ee%
where %
\be\label{f0}%
f_0=1+ \frac{4(q_2q_3+q_2q_4+q_3q_4)}{L^2}%
\ee%
and $q_i, \ \mu_c, \ M; \rho,\ \tau,\ \varphi,\ \psi_i$ are fixed
in the $\epsilon\rightarrow0$ limit. Note also that in this limit
\begin{equation}
f=f_0\ (1-\frac{M}{\rho^2}).
\end{equation}

Performing the above limit on \eqref{11-metric-original} will
result in the following metric
\begin{equation}\label{near-ext-metric}%
{ds_2}=\mu_1^{4/3}\  \left(R_A^2 ds_{BTZ}^2+R^2_S
d\Omega_2^2\right)+ \frac{1}{\mu_1^{2/3}}\ ds_{{\cal M}_6}^2
\end{equation}
where
\begin{equation}\label{near-ext-BTZ}%
ds_{BTZ}^2=-(\rho^2-M)\
d\tau^2+\frac{d\rho^2}{\rho^2-M}+\rho^2d\varphi^2,
\end{equation}%
and
\begin{equation}\label{near-ext-M6}%
ds_{{\cal M}_6}^2=\frac{L^2}{R_S}\ \sum_{i=2,3,4}
q_i(d\mu_i^2+\mu_i^2d\psi_i^2).
\end{equation}%
with
\begin{equation}\label{Rs-Ra-ext}%
R_S^3=q_2 q_3 q_4=\frac{\mu_c L^2}{4}, \qquad
R_A^2=\frac{4}{{f_0}}\ R_S^2.
\end{equation}%

For $M>0$, the metric \eqref{near-ext-BTZ} \emph{locally}
describes a stationary BTZ black hole with mass $M$. For $M<0$ we
have a conical space with a deficit angle $2\pi(1-\epsilon^2M)$.
Also note that the range of the angle $\varphi$ in $AdS_3$ is
$[0,2\pi\epsilon]$. Nonetheless, the causal boundary of this
locally $AdS_3$ space is still $R\times S^1$ \cite{FGMS-1}.

The metric \eqref{near-ext-metric} is a solution to
eleven-dimensional supergravity with the three-form%
\be\label{near-ext-3form}%
C_3=-\frac{L^2}{2} \sum_{i=2,3,4} \tilde q_i\ \mu_i^2
d\psi_i\wedge
d^2\Omega_2%
\ee%
where in the near-horizon far-from-BPS limit%
\be\label{tildeq-q}%
\tilde q_i^2=q_i(q_i+\mu_c)\ , \qquad i=2,3,4.%
\ee%

\subsection{Perturbative addition of the fourth charge}

The black hole solution we have discussed so far has been a three
charge one. When the non-extremality parameter is small, the
near-horizon limit of these black holes resulted in decoupled
geometries having an $AdS_3\times S^2$ factor as a subspace. In this
section we extend our analysis of these black holes by turning on
the fourth charge. We require, for both of the near-BPS and
far-from-BPS cases, that this charge to be much smaller than the
other three, $q_1\ll q_2,\ q_3,\ q_4$, so that it can be considered
as a perturbation to the previous case. One expects that the
near-horizon limit of such four-charge near extremal black holes
results in similar decoupled geometries as above but with an
additional angular momentum in the $AdS_3$ factor to obtain a
rotating BTZ black hole. In the following we will show that this is
in fact the case for both the near and far-from-BPS
cases.\footnote{One can study near-horizon limit of a generic
extremal four-charge black hole (when all four charges are of the
same order). In the near-horizon limit a generic four-dimensional
four-charge black hole leads to $AdS_2 \times S^2$ \cite{Cucu1}.
This near-horizon limit and application of entropy function to study
the entropy of these four-charge extremal black holes has been
carried out in \cite{Morales:2006gm}.}

\subsubsection{The near-horizon, near-BPS case}\label{near-BPS-rot-section}%

To extend  the near-horizon limit to include the fourth charge $q_1$, we supplement the limit \eqref{near-limit-BPS-1} or
\eqref{near-BPS-limit-2} with
\begin{equation}\label{near-BPS-J-limit}
q_1=\epsilon^3 \hat{q}_1, \qquad r=\epsilon^3(\hat{\mu}_c
\rho^2-\hat{q}_1),
\end{equation}
while keeping $\hat q_1$ fixed. As discussed at the end of section
2, when the fourth charge is also turned on one can extend the
geometry past $r=0$, to $r\geq -q_1$. The shift in the scaling of
$r$ in \eqref{near-BPS-J-limit} is a reflection of this fact. In the
limit $\epsilon\rightarrow 0$
\begin{equation}
f=1-\frac{\hat{\mu}_c \gamma}{\hat{\mu}_c
\rho^2-\hat{q}_1}+\frac{\hat{\mu}_c\hat{q}_1}{(\hat{\mu}_c
\rho^2-\hat{q}_1)^2}, \qquad a_1=-\frac{{\cal J}}{2\rho^2}\ ,
\end{equation}
where $\gamma$ is defined in \eqref{gamma} and
\begin{equation}\label{cal-j-BPS}
{\cal
J}\equiv2\frac{\sqrt{\hat{q}_1(\hat{q}_1+\hat{\mu})}}{\hat{\mu}_c}\
.
\end{equation}
Performing the limit on \eqref{11-metric-original} we find the
metric in \eqref{nearBPSmetric} with the same expressions for
$R_A$ and $R_S$, but the $ds^2_{BTZ}$ is now replaced with
\begin{equation}\label{near-BPS-Rot-BTZ}
ds_{AdS}^2=-\frac{F(\rho)}{\rho^2}\ dt^2+
\frac{\rho^2}{F(\rho)} d\rho^2+\rho^2(d\phi-\frac{{\cal J}}{2\rho^2}\ dt)^2.
\end{equation}
where
\begin{equation}\label{F-rho-BPS}%
F(\rho)=\rho^4-(\gamma+2\frac{\hat{q}_1}{\hat{\mu}_c})\rho^2+\frac{{\cal J}^2}{4}
\end{equation}
The metric \eqref{near-BPS-Rot-BTZ} describes a rotating BTZ with
(see Appendix A of \cite{FGMS-1} for our conventions)
\begin{equation}\label{BPS-M,J}
M_{BTZ}=\gamma+2\frac{\hat{q}_1}{\hat{\mu}_c}=\frac{\hat\mu+2\hat
q_1}{\hat\mu_c}-1, \qquad J_{BTZ}={\cal J}.
\end{equation}

\subsubsection{The near-horizon,  far-from-BPS case}\label{Rot-BTZ-NExt-limit-section}%
 Again the only ingredient we need to add to \eqref{near-ext-limit} is the scaling of the fourth charge,
\begin{equation}\label{near-ext-rot-limit}
q_1=\epsilon^4 \hat{q}_1,
\end{equation}
while keeping the rest of the scalings unchanged. In the limit
\begin{equation}
f=f_0(1-\frac{M}{\rho^2}+\frac{J^2}{4\rho^4}), \qquad a_1=-\sqrt{f_0}\ \frac{J}{2\rho^2},
\end{equation}
where $M$, as defined in \eqref{near-ext-limit}, is
$M=\frac{\mu-\mu_c}{\epsilon^2\mu_c}$ and
\begin{equation}\label{j}
J\equiv 2\left(\frac{\hat{q}_1f_0}{\mu_c}\right)^{1/2}.
\end{equation}

Performing the limit on \eqref{11-metric-original}, we end up with
the metric \eqref{near-ext-metric} but $ds^2_{BTZ}$ replaced with
\begin{equation}\label{near-ext-rot-BTZ}
ds_{AdS}^2=-\frac{G(\rho)}{\rho^2}\ d\tau^2+\frac{\rho^2}{G(\rho)}\ d\rho^2+\rho^2(d\varphi^2-\frac{J}{2\rho^2}\ d\tau)^2,
\end{equation}
where
\begin{equation}\label{G-rho-Extremal}%
G(\rho)=\rho^4-M \rho^2+\frac{J^2}{4}.
\end{equation}
The metric \eqref{near-ext-rot-BTZ} describes a rotating BTZ with
\begin{equation}\label{ext-M,J}
M_{BTZ}=M, \qquad J_{BTZ}=J\ .
\end{equation}

Similar to the non-rotating case of section 3.2, the metric obtained
in this case is also a solution to eleven-dimensional supergravity
with the three-form field given in \eqref{near-ext-3form}. The
explicit verification of this point will be given in the Appendix.

\section{The  $5d$ Description of BTZ$\times S^2$}
\label{Section4-5d-gravity}

In this section we show that the BTZ$\times S^2$ geometries of
previous section can be realized as solutions to five-dimensional
supergravities. The near-BPS case is a solution to \emph{ungauged}
STU model and the far-from-BPS case is a solution to \emph{gauged}
$U(1)^3$ supergravity. We also show that the former can be obtained
as near-horizon limit of magnetically charged string solutions of
the the STU model, first constructed in \cite{Chamseddine:1999qs}.

\subsection{The near-BPS case}

The action of 5d ungauged STU model \cite{Chou:1997ba,Sabra:1997yd}
is given by
\begin{equation}\label{STU-ungauged-action}
\begin{split}
    S_{ungauged}=
    \frac{1}{16\pi G_N^{(5)}}\int  dx^5\sqrt{-g^{(5)}}
    &\biggl(R^{(5)}-\sum_{i=1,2,3}\bigl(\frac12\,(X^i)^{-2}\partial_{\mu}X^i\,\partial^\mu X^i
     \\
    & +\frac14(X^i)^{-2}
    F^{i}_{\mu\nu}\,F^{i\,\mu\nu}\bigr)\biggr)+
    \frac{1}{4}\epsilon^{\mu\nu\rho\sigma\lambda}F^{1}_{\mu\nu}F^{2}_{\rho\sigma}
    A^{3}_{\lambda}\ ,
\end{split}
\end{equation}
where scalars $X^i$ obey the constraint
\begin{equation}\label{constraint-X-I}
    X^1X^2X^3=1\ .
\end{equation}
This action is a truncation  of the general ${\cal N}=2$, $d=5$
supergravity \cite{Gunaydin:1983bi} obtained from reduction of
$11d$ supergravity on Calabi-Yau threefold \cite{Cadavid:1995bk,
Papadopoulos:1995da}. Moreover, \eqref{STU-ungauged-action} can be
obtained by compactification of heterotic string theory on
$K_3\times S^1$ \cite{Antoniadis:1995vz}.

We seek a string solution of this theory which has
BTZ$\times S^2$ geometry. Our ansatz for the field configuration of
this solution is
\begin{subequations}\label{ansatz-bps-fields}
\begin{align}
ds^2=R^2_A\Big(-f(\rho)d\tau^2+\frac{d\rho^2}{f(\rho)}&+\rho^2d\phi_1^2\Big)+R^2_S\Big(d\theta^2+sin^2\theta d\phi^2\Big)\\
X^i&=u^i\\
F^i_{\theta\phi}&=p^i\ sin\theta
\end{align}
\end{subequations}
where the function $f(\rho)$ and constants $R_A,\ R_S, u_i$ are
determined in terms of the magnetic charges $p_i$ using the
equations of motion.

It is straightforward to check that the above ansatz is a solution
to the STU model \eqref{STU-ungauged-action} if %
\be\label{f-rho-ansatz}%
f(\rho)=\rho^2-M+\frac{J^2}{4\rho^2}%
\ee%
and when%
\begin{equation}\label{solution-bps}%
    R_S=\frac{R_A}{2}=\frac{p^i}{u^i}\ ,\quad i=1,2,3.
\end{equation}%
The constraint  \eqref{constraint-X-I} implies $u^1u^2u^3=1$ and hence %
 \be\label{Radius-pi}%
R_S^3= p^1p^2p^3.%
\ee%
Note that $M$ and $J$ in \eqref{f-rho-ansatz} are independent of the $p^i$'s. That is, the $BTZ\times S^2$ geometry we
obtained as a result of taking the near-horizon near-BPS limit is
a solution to STU model once we rename magnetic charges as%
\begin{eqnarray}\label{pi-qi}%
 p^i&=&(\mu_1^0)^{2/3}\hat q_i ~~\text{for} ~~ \mu_1 \sim \mu_1^0, \\ \nonumber
p^i&=& \hat q_i ~~ \text{for}~~\mu_1 \sim 1.
\end{eqnarray}%
In the conventions of \cite{Chamseddine:1999qs, Sabra-Klemm}, the
above BTZ$\times S^2$ solutions are characterized by the
``magnetic central charge'' $Z=\frac13 u^i p_i=R^3_S$.

As discussed in \cite{AdS3-AdS2} it is possible to obtain the same
BTZ$\times S^2$ solution as the near-horizon limit of ``near-BPS''
magnetically charged string solutions, which are charged under all
three $U(1)$ fields of the STU model.

From the $11d$ viewpoint these black strings can be obtained as
geometries corresponding to three stacks of intersecting M5-branes
wrapping holomorphic four-cycles of a CY threefold
\cite{Behrndt:1996he, Vijay-Larsen} whose near-horizon limit
coincides with the $11d$ solutions we obtained after the
near-horizon, near-BPS limit \eqref{nearBPSmetric}. In our case too
the $AdS_3\times S^2$ is coming as near-horizon limit of specific
intersecting M5-branes, but spherical (rather than flat) M5-branes
in the $AdS_4\times S^7$ background. Note, however, that the BTZ and
the $AdS_3$ factor we obtain in our limit is already in
\emph{global} coordinates, while those obtained in
\cite{Chamseddine:1999qs,Behrndt:1996he,Vijay-Larsen} have $AdS_3$
in the Poincare patch. From the $AdS_3\times S^2\times {\cal C}_6$
geometry \eqref{nearBPSmetric} one can read off the $5d$ Newton
constant in terms of the eleven-dimensional one,%
\be\label{STU-GN}%
G^{(5)}_N=\frac{G^{(11)}_N}{V_{{\cal C}_6}}\cdot
\epsilon^{-3}=\frac{16\pi^7 l_p^9}{{(2\pi)^3} L^6 \cdot
\frac{\mu_2^0\mu_3^0\mu_4^0}{(\mu_1^0)^2}}\cdot \epsilon^{-3}
=\frac{\pi }{64\sqrt{2}}\ L^3\ (N\epsilon^2)^{-3/2}\
\frac{(\mu_1^0)^2}{\mu_2^0\mu_3^0\mu_4^0}
. \ee%
Note that, in the above, after the reduction on the ${\cal C}_6$,
to remove the $\epsilon$ factor appearing in front of the metric,
we have rescaled the resulting $5d$ metric by a factor of
$\epsilon$.

As discussed in \cite{Chamseddine:1999qs, Klemm:2000nj,Sabra-Klemm}
the rotating BTZ$\times S^2$ solutions for $M=J\ge0$, where we have
$AdS_3\times S^2$ geometry, are half BPS, meaning that they preserve
4 out of 8 supersymmetries of the ${\cal N}=2$, $5d$ theory.  The
supersymmetry of the magnetically charged black strings is half of
supersymmetry of the rotating BTZ$\times S^2$ obtained after the
near-horizon limit.

A few comments about the near-horizon near-BPS limit that we have
presented above are in order here. While we started from a $4d$
\emph{electrically} charged black hole solution, after the limit we
ended up with a $5d$ \emph{magnetically} charged solution. In the
literature \cite{AdS3-AdS2}, there exists a duality between
electrically and magnetically charged solutions to $5d$ STU model.
This duality relates electric $AdS_2 \times S^3$ solutions to the
magnetic $AdS_3 \times S^2$ and was established in \cite{AdS3-AdS2}
by reducing the solutions to $4d$, more precisely to solutions of
$4d$ ungauged STU model, where the duality is an electric-magnetic
S-duality.

This duality between electrically and magnetically charged solutions
may be relevant for us. Although the electrically charged $4d$ black
hole is a solution to the gauged SUGRA,  in the near-horizon,
near-BPS limit that we take, the $X_i$ (\emph{cf.} \eqref{4dBH}) are
scaled in such a way that the potential term in the gauged SUGRA
vanishes and hence it  reduces to an ungauged theory
 action.

\subsection{The far-from-BPS case}

The action for the $5d$ gauged $U(1)^3$ supergravity is
\cite{Gunaydin:1984ak} (for a review \emph{e.g.} see \cite{Duff:1999rk})%
\begin{equation}\label{gauged-action}
\begin{split}
    S_{gauged}=\frac{1}{16\pi G_N^{(5)}}\int  dx^5\sqrt{-g^{(5)}}
    &\biggl(R^{(5)}-\sum_{i=1,2,3}\bigl(\frac12\,(X^i)^{-2}\partial_{\mu}X^i\,\partial^\mu X^i-
    \frac{4}{L^2} {(X^i)}^{-1} \\
    & +\frac14(X^i)^{-2}
    F^{i}_{\mu\nu}\,F^{i\,\mu\nu}\bigr)\biggr)+
    \frac{1}{4}\epsilon^{\mu\nu\rho\sigma\lambda}F^{1}_{\mu\nu}F^{2}_{\rho\sigma}
    A^{3}_{\lambda}\ ,
\end{split}
\end{equation}
where $X$'s are subject to the constraint \eqref{constraint-X-I}. The
above action is  the same as \eqref{STU-ungauged-action} but with
the extra potential term for $X^i$'s. This potential has appeared
as a result of the gauging of $U(1)^3$ in the supergravity.

To check that the BTZ$\times S^2$ geometry we obtained from the
near-horizon far-from-BPS limit of section 3.2 is a solution to the
above theory, we plug the ans\"atz \eqref{ansatz-bps-fields} into
the action and solve for $R_A$ and
$R_S$ in terms of the magnetic charges $p^i$'s. We then get %
\begin{subequations}\label{exremal-as-5d-solution}
\begin{align}
R^2_A&=\frac{4L^2(Q_1Q_2Q_3)^\frac23}{L^2+4(Q_1Q_2+Q_1Q_3+Q_2Q_3)}\\
R^2_S&=(Q_1Q_2Q_3)^\frac23\\
u^i&=\frac{Q_i}{(Q_1Q_2Q_3)^\frac13}
\end{align}
\end{subequations}
where $Q^i$'s are given in terms of $p^i$'s through the relations
\begin{equation}\label{def-Q-i}
    p^i=\sqrt{Q_i\left(Q_i+\frac{4Q_1Q_2Q_3}{L^2}\right)}\qquad\qquad(i=1,2,3).
\end{equation}%
It is readily seen that the above is the same $AdS_3\times S^2$
geometry of section 3.2 once we identify $Q^i$'s with the $q^i$'s
there.

To read the $5d$ Newton constant in this case, we recall the
decoupled metric \eqref{near-ext-metric} and do the reduction of
the $11d$ SUGRA over ${\cal M}_6$ \eqref{near-ext-M6}:
\be\label{gaugedU(1)3-GN}%
G^{(5)}_N=\frac{G^{(11)}_N}{V_{{\cal M}_6}}=\frac{16\pi^7
l_p^9}{\frac{\pi^3}{6} L^6}=\frac{3\pi }{4\sqrt{2}} L^3\ N^{-3/2}\
, \ee%
where the $V_{{\cal M}_6}$ is the volume of the six-dimensional
manifold ${\cal M}_6$ \eqref{near-ext-M6}. Note that the range of
$\mu_i$'s in ${\cal M}_6$ is such that $\mu_2^2+\mu_3^2+\mu_4^2=1$
and all of them are non-negative.\footnote{This is in contrast
with the original eleven-dimensional metric in which
$\mu_2^2+\mu_3^2+\mu_4^2=1-\mu_1^2$.}

It is straightforward to examine  whether the above $AdS_3\times
S^2$ geometry is a supersymmetric  solution to the $5d$
gauged SUGRA by checking the Killing spinor equations. It turns
out that the above solution does not preserve any supersymmetry.
This is consistent with the results of \cite{Sabra-Klemm, Cucu1,
Cucu-thesis} and \cite{Gut-Sabra} where the classification of all
BPS solutions of this $5d$ gauged SUGRA has been carried out. We
would also like to comment that, although we expect it to exist,
the magnetic string solution to the gauged SUGRA which leads to
the above $AdS_3\times S^2$ geometry in the near-horizon limit has
not yet been constructed.

\section{Entropy Analysis of the $4d$ and $3d$ Black Holes}\label{Section5-entropy}

In this section, we first compute the entropy of the
four-dimensional near extremal black hole for both the near-BPS and
far-from-BPS cases. We do this for a four-charge black hole, one of
the charges being much smaller than the rest. We will then compute
the entropy of the rotating BTZ black hole which is a subspace of
the decoupled geometry after the near-horizon limit. We find that
the results coincide \emph{i.e.} the four and three dimensional
black holes, in the limits of our interest, produce the same value
for the entropy. This  provides supportive evidence for the fact
that our limits are indeed  decoupling limits.

\subsection{$4d$ black hole entropy}
As mentioned earlier, the four-dimensional black holes discussed
above are solutions to four-dimensional $U(1)^4$ gauged SUGRA with
the metric given in \eqref{4dBH}. The position of the horizon
$r_h$ is determined by the zeroes of $f/H^{1/2}$ and
\begin{equation}\label{4d-BH-entropy}%
S^{(4)}_{B.H.}=\frac{A_h}{4G^{(4)}_N},\qquad A_h=4\pi r_h^2
(H_1H_2H_3H_4)^{1/2}|_{r=r_h}\ .
\end{equation}
Using \eqref{GN-4-11} that is%
\be\label{4d-BH-entropy-N}%
S_{BH}=\frac{2\sqrt2}{3}\ N^{3/2}\cdot \frac{A_h}{L^2}\ . %
\ee%

As we see for generic horizon areas of order $L^2$ the entropy
scales as $N^{3/2}$. The three-charge extremal black holes have
vanishing horizon size. However the near extremal solutions have
a non-zero horizon size, where their horizon areas scale to zero with some
power of $\epsilon$. Moreover,  as discussed in section 2, the
$r=0$ is the curvature singularity of the three-charge black
holes. Therefore, it is necessary to make sure that in our limit
we can still trust (classical) gravity description. In order this
we should make sure that%
\be\label{large-entropy}%
S_{BH}\gg 1, \ee %
and that all the curvature invariants of the ``decoupled''
geometries remain small (in the relevant Planck units). The
condition \eqref{large-entropy} can only be fulfilled  when together
with taking $\epsilon\to 0$ we also take the large $N$ limit in an
appropriate rate.

\subsubsection{The near-BPS limit}\label{near-BPS-entropy-section}

The horizon radius is the larger root of function $f/H^{1/2}$,
which turns out to be same as the larger root of $F(\rho)$,
\eqref{F-rho-BPS}%
\begin{equation}
r_h+q_1=\frac{\hat{\mu}_c}{4}\left(\sqrt{\frac{\mu+2{q}_1}{{\mu}_c}-{\cal
J}-1}+\sqrt{\frac{\mu+2{q}_1}{{\mu}_c}+{\cal J}-1}\right)^2\
\epsilon^3,
\end{equation}
and the horizon area is
\begin{equation}
A_h=2\pi L \hat{\mu}_c^{1/2} (r_h+q_1)^{1/2} \epsilon^{3/2},
\end{equation}
where ${\cal J}$ is defined in \eqref{cal-j-BPS}. The entropy is  found to be
\begin{equation}\label{entropy-near-BPS-4d}
S^{Near-BPS}_{BH}=\frac{2\sqrt2\pi}{3}\ \frac{\hat{\mu}_c}{L}\
\left(\sqrt{\frac{\mu+2{q}_1}{{\mu}_c}-{\cal
J}-1}+\sqrt{\frac{\mu+2{q}_1}{{\mu}_c}+{\cal J}-1}\right)
\left(N\epsilon^2\right)^{\frac32}\ .
\end{equation}

As the above Bekenstein-Hawking entropy only makes sense for
classical black holes, we should make sure that
\eqref{large-entropy} is fulfilled. This is only possible if
together with $\epsilon$ we scale $N\sim \epsilon^{-\alpha}$,
$\alpha \geq 2$. The components of the curvature tensor for the
decoupled geometry all scale as $\epsilon^{-2}$ (in units of $L$)
which in the limit remain large. Noting the factor of $\epsilon^2$
in front of the decoupled metric components \eqref{nearBPSmetric}
and that eleven-dimensional Planck length should be the shortest
length,%
\be\label{N-epsilon-BPS}%
\epsilon\sim l_p/L\ \Rightarrow N\sim \epsilon^{-6}\to\infty\ ,%
 \ee%
where we have used \eqref{GN-4-11} and that $L$ is fixed. This
choice of \eqref{N-epsilon-BPS} will become more clear in the next
section. With the above scaling $S^{Near-BPS}_{BH}\sim
N\to\infty$. 

As reviewed in section 2, the three charge black holes at the extremal point have vanishing horizon and 
hence a naked singularity. One may then ask if the near horizon limit leads to a valid description within the supergravity approximation. As we argued above, in the near extremal limits we have discussed certain $l_p\to 0$ limit is also taken such that the entropy of the BTZ black hole is non-vanishing. Moreover, we note that 
in the same limit the radii of the $AdS_3$ and $S^2$ factors (in units of $11d$ Planck length) remains finite and large. Therefore the near horizon geometry is within the regime of the validity of supergravity.

\subsubsection{The far-from-BPS
case}\label{near-ext-entropy-section}

In the far-from-BPS scaling the horizon radius is the larger root of
$f/H^{1/2}$, which turns out to be the same as the larger root of
$G(\rho)$ \eqref{G-rho-Extremal},
\be\label{rh-extremal}%
r_h =\frac{\mu_c}{4f_0}\left(\sqrt{M-J}+\sqrt{M+J}\right)^2
\epsilon^2,
\ee
and the horizon area is
\be
A_h =2\pi L \mu_c^{1/2} r_h^{1/2}=\pi \frac{\mu_c
L}{\sqrt{f_0}} \left(\sqrt{M-J}+\sqrt{M+J}\right) \epsilon,
\ee%
where $M$ and $J$ are defined in \eqref{near-ext-limit} and
\eqref{j} respectively. The Bekenstein-Hawking entropy is found to be
\begin{equation}\label{entropy-near-ext-4d}%
S^{\textrm{far-from-BPS}}_{BH}=\frac{2\sqrt2 \pi}{3}\
\frac{\mu_c}{\sqrt{f_0}L}\  (N^{3/2}\epsilon)
\left(\sqrt{M-J}+\sqrt{M+J}\right).
\end{equation}%

To ensure validity of our classical treatments and
\eqref{large-entropy} we need to scale $N\sim \epsilon^{-\beta}\to
\infty$, $\beta\geq 3/2$ as we take $\epsilon\to 0$. Arguments of
next section and the $3d$ CFT analysis specifies $\beta=2$, that
is
\be\label{N-epsilon-extremal}%
N\sim \epsilon^{-2}\ ,\qquad \epsilon\sim
\left(\frac{l_p}{L}\right)^{1/3}.
\ee%
With this choice $S^{textrm{far-from-BPS}}_{BH}\sim N$. It is
notable that the scaling of entropy with $N$ in both of the the
near-BPS and far-from-BPS cases are the same. This is in accord with
similar results  for five-dimensional two-charge black holes
\cite{FGMS-1}. Note also that the radii of $AdS_3$ and $S^2$ and the cycles in the ${\cal M}_6$ remain finite and large in $11d$ Planck units. \footnote{Note also that the geometry \eqref{near-ext-metric} has a curvature singularity at $\mu_1=0$.  This is not going to invalidate our near horizon limit and validity of supergravity because this singularity is an artifact of the near horizon limit \eqref{near-ext-limit}. At $\mu_1=0$ one should revisit the limiting procedure more carefully. Doing so, it is readily seen that the singularity is resolved.}

\subsection{$3d$ rotating BTZ entropy}

As discussed in the near-horizon limit for both of the near-BPS and
far-from-BPS cases we obtain a rotating BTZ black hole, \emph{cf.}
\eqref{near-BPS-Rot-BTZ} and \eqref{near-ext-rot-BTZ}. In this
sub-section, we compute the Bekenstein-Hawking entropy of these
three-dimensional black holes and compare it to the entropy of
four-dimensional black holes computed in the previous subsections.
As we will see they are equal. But first, we need to compute the
corresponding three-dimensional Newton constant $G_N^{(3)}$. As
discussed in section 4, both the near-BPS and far-from-BPS rotating
BTZ$\times S^2$ geometries are solutions to five-dimensional
supergravities  both of which have the same Newton constant
\eqref{STU-GN}, \eqref{gaugedU(1)3-GN}. The rotating BTZ black hole
is then the solution to the three-dimensional gravity obtained from
the reduction of (either of) the five-dimensional gravity theories
on the $S^2$ of radius $R_S$, therefore
\begin{equation}\label{GN-3d}
G_N^{(3)}=\frac{G^{(5)}_N}{4\pi R^2_S}\ .
\end{equation}
The Bekenstein-Hawking area-law for a rotating BTZ of angular
momentum $J$ and mass $M$ is (\emph{e.g.} see \cite{Justin-Mandal})
\begin{equation}\label{entropy-BTZ}%
S_{BTZ}=\frac{\pi}{2G_N^{(3)}} R_A
\left(\sqrt{M_{BTZ}-J_{BTZ}}+\sqrt{M_{BTZ}+J_{BTZ}}\right)\ ,
\end{equation}
where $R_A$ is the $AdS$ radius. Next we consider the  near-BPS and
far-from-BPS cases separately.

\subsubsection{The near-BPS case}

In taking the near-BPS limit  we are focusing on fixed values for
$\theta_i$ and hence each near-horizon geometry describes a
 ``{\emph{strip}}" of the original four-dimensional black
hole, similarly to the five-dimensional case discussed in
\cite{Balasubramanian:2007bs, FGMS-1}. The entropy of the
four-dimensional black hole is hence expected to be distributed
among these strips each corresponding to a rotating BTZ$\times S^2$.
The mass and angular momentum of all of these BTZ's, in units of the
corresponding $AdS_3$ radius, are equal. However, these BTZ'z have
different $AdS_3$ radii ({\it cf.} \eqref{Rs-mu-neq1}). Moreover,
the corresponding three-dimensional Newton constant is different for
each of them, depending on the value of $\mu_i^0$. The entropy of
each strip is then%
\be\label{strip-entropy}%
dS_{strip}=2\pi R_A\ \cdot \frac{(2\pi)^3 L^6\cdot 4\pi
R^2_S}{4G_N^{(11)}}\
\left(\sqrt{M_{BTZ}+J_{BTZ}}+\sqrt{M_{BTZ}-J_{BTZ}}\right)
\epsilon^3\ \frac{1}{8}d\hat{\mu}_2^2\ d\hat{\mu}_3^2\ d\hat{\mu}_4^2, %
\ee%
where $R_A$ and $R_S$ are respectively given in \eqref{Rs-Ra-BPS}
and \eqref{Rs-mu=1}, \eqref{Rs-mu-neq1}. Noting that
\begin{equation}
\int_{\mu_2^2+\mu_3^2+\mu_4^2=1} d\hat{\mu}_2^2\ d\hat{\mu}_3^2\
d\hat{\mu}_4^2=\frac{1}{6},
\end{equation}
the total entropy becomes
\begin{equation}\label{entropy-near-BPS-3d}
S_{BTZ}=\frac{2\sqrt{2}\pi}{3} \left(N\epsilon^2\right)^{3/2}\
\frac{\hat{\mu}_c}{L}\
\left(\sqrt{M_{BTZ}-J_{BTZ}}+\sqrt{M_{BTZ}+J_{BTZ}}\right),
\end{equation}
where $M_{BTZ}$ and $J_{BTZ}$ are given by \eqref{BPS-M,J}. As we
see there is a perfect matching between the four-dimensional
entropy \eqref{entropy-near-BPS-4d} and the collection of the
``strip-wise'' three-dimensional rotating BTZ black holes
\eqref{entropy-near-BPS-3d}.

\subsubsection{The far-from-BPS case}

The three-dimensional Newton constant \eqref{GN-3d} is%
\be\label{GN-Next}%
G_N^{(3)}=\frac{3}{16\sqrt2}\ \frac{L^3}{R_S^2}\ N^{-\frac32}, %
\ee%
where $R_S$ is given by \eqref{Rs-Ra-ext}. Noting that the angular
variable in BTZ, $\varphi$ ranges over $[0,2\pi\epsilon]$, the
entropy is then%
\begin{equation}\label{entropy-near-ext-3d}%
S_{BTZ}=\frac{2\sqrt2\pi}{3}\ \left(N^{3/2}\epsilon\right)
\frac{\mu_c}{L}\
\left(\sqrt{M_{BTZ}-J_{BTZ}}+\sqrt{M_{BTZ}+J_{BTZ}}\right)\ ,
\end{equation}
where $M_{BTZ}$ and $J_{BTZ}$ are given in \eqref{ext-M,J}. Again
we see a perfect matching between the three and four dimensional
entropies.

\section{Dual Field Theory Descriptions}\label{Section-dual-CFTs}%

In this section we study the near-horizon decoupling limits from
the dual (conformal) field theory viewpoints. First we give the
$3d$ description, motivated by the fact that the original geometry
is a four-dimensional black hole in the $AdS_4$ background. Next,
we focus on the $2d$ dual field theory description arising from
the appearance of $AdS_3$ factors in the decoupled geometries.

\subsection{$3d$ CFT description}

According to the $AdS_4/CFT_3$ duality \cite{MAGOO} there is a
one-to-one correspondence between the four-dimensional black holes,
as deformations about the $AdS_4\times S^7$ geometry and certain
sectors of the $3d$ CFT. We choose the $AdS_4\times S^7$ background
to correspond to the near-horizon limit of $N$ coincident M2-branes.
The operators of this $3d$ CFT are specified by $SO(3,2)\times
SO(8)$ quantum numbers. In our case the five parameters of the black
hole geometry, $q_i$ and $\mu$, are mapped to the four R-charges
$J_i$ and the engineering
dimension of the operators $\Delta$ as %
\be\label{Delta,J}
\begin{split}%
\Delta &=L\cdot M_{ADM}=\frac{4\sqrt2}{3}\ {N^{3/2}}\
(2\mu+q_1+q_2+q_3+q_4)/L\ ,\cr J_i &=\frac{L}{2G_N^{(4)}} \tilde
q_i=\frac{4\sqrt2}{3}\ N^{3/2}\ \frac{\tilde q_i}{L}\ .
\end{split}\ee%
where $M_{ADM}$ is given in \eqref{4d-ADM-mass}. Operators of
interest to us are singlets of $SO(3)\in SO(3,2)$.

As discussed in the previous section, in both of the near-BPS and
far-from-BPS limits we are taking $N\to \infty$, and $\Delta$ and
$J_i$ are hence becoming large. Similarly to the case of two-charge
five-dimensional black holes of \cite{FGMS-1}, we search for a
``BMN-type'' sector in the $3d$ CFT whose dynamics is decoupled from
the rest of the theory.

\subsubsection{The near-horizon near-BPS limit, ${\cal N}=8$ $3d$ CFT
description}\label{NBPS-SYM-section}

In the near-BPS limit together with some of the coordinates we also
scale $\mu\sim \epsilon^3$ and  $q_i\sim \epsilon$. As discussed in
section \ref{near-BPS-entropy-section}   we need to also scale
$N\sim \epsilon^{-6}$, which we choose
 \be\label{eps-lambda}
\epsilon=\frac{1}{\sqrt2} N^{-1/6}\ .
\ee%
Therefore, in this limit, for the three-charge case $\Delta$ and
$J_i$ of the
operators scale as:%
\be\label{Delta-J-BPS-limit}
\begin{split}
\Delta &=\frac{4\sqrt2}{3}\ N^{3/2}\epsilon\ (\hat q_2+\hat q_3+
\hat q_4+{\cal O}(\epsilon^2))/L\sim N^{4/3}\to\infty\\ J_i&=
\frac{4\sqrt2}{3}\ N^{3/2}\epsilon\ (\hat q_i+ {\cal
O}(\epsilon))/L\sim N^{4/3}\ \qquad i=2,3,4\ .
\end{split}
\ee%
That is, the sector of the ${\cal N}=8$ $3d$ CFT corresponding to
M-theory on the geometries in question have large scaling dimension
and $R$-charge, $\Delta\sim J_i\sim N^{4/3}$. In the same spirit as
the BMN limit \cite{BMN,FGMS-1}, one can find certain combinations
of $\Delta$ and $J_i$ which are finite and describe physics of the
operators after the limit. To find these combinations we recall the
way the limit was taken
\eqref{near-limit-BPS-1}, %
\begin{subequations}\label{Delta-J-BPS-1}%
\begin{align}%
iL\frac{\partial}{\partial \tau}&=iL\frac{\partial}{\partial
t}+i\sum_{i=2,3,4} \frac{\partial}{\partial \phi_i} =\Delta-\sum_{i=2,3,4} J_i\\
-i\frac{\partial}{\partial \psi_i}&=-i\frac{\partial}{\partial
\phi_i}=J_i\ .
\end{align}%
\end{subequations}%
For the limit \eqref{near-BPS-limit-2}, (\ref{Delta-J-BPS-1}b)
should be replaced with $-i\frac{\partial}{\partial
\psi_i}=-i\epsilon \frac{\partial}{\partial \phi_i}=\epsilon J_i.$
 Up to leading order we have
\be\label{Delta-J-BPS-2}%
\begin{split}%
\Delta-\sum_{i=2,3,4} J_i&=\frac{2\sqrt2}{3}\ N^{3/2} \epsilon^3\
\frac{\hat\mu}{L}\ , \cr J_i&= \frac{4\sqrt2}{3}\ N^{3/2}\epsilon\
\frac{\hat q_i}{L},\ \qquad i=2,3,4\ .
\end{split}%
\ee%
As we see $\Delta-\sum J_i$ scales as $N^{3/2}\cdot
N^{-1/2}=N\to\infty$, while $J_i\sim N^{4/3}$ and therefore the
``BPS-deviation-parameter'' \cite{FGMS-1}
\be\label{BPS-deviation} %
\eta_i\equiv \frac{\Delta-\sum_i
J_i}{J_i}\sim \epsilon^2\sim N^{-1/3}\to 0\ , %
\ee%
and hence we are dealing with an ``almost-BPS'' sector.\footnote{It
is instructive to make parallels with the BMN sector \cite{BMN}. In
the BMN sector of the $3d$ CFT we are dealing with operators with
\[
\Delta\sim J\sim N^{1/3},\qquad {\rm while}\qquad \Delta-J\ =finite,
\]
implying that, similar to our case, $\eta_{BMN}\sim N^{-1/3}\to 0$.
As we see the $\eta$ parameter for our case and the case of BMN
scale in the same way.} Moreover, $\Delta-\sum J_i$ is linearly
proportional to the non-extremality parameter $\hat\mu$. It is also
notable that $S_{BH}$ \eqref{entropy-near-BPS-4d} scales the same as
$\Delta-\sum J_i$.

In sum, the sector we are dealing with is  composed of ``almost 1/8 BPS'' operators of  the ${\cal N}=8$ $3d$ CFT, with%
\be\label{SYM-observables}%
\begin{split}
\Delta\sim & J_i\sim N^{4/3} , \\
\frac{J_i}{N^{4/3}}\equiv \frac{4{\hat q}_i}{3L}=fixed &, \qquad
(\Delta-\sum_{i=2,3,4} J_i)\cdot \frac{1}{N}=\frac{\hat
\mu}{3L}=fixed.
\end{split}%
\ee%
The  dimensionless physical quantities that describe this sector
are therefore $\hat q_i/L,\ \hat \mu/L$.

Here we are dealing with a system of intersecting multi M5-brane
giants. The ``number of giants'' in each stack
\eqref{five-brane-number} in the near-BPS,
near-horizon limit is %
\be\label{giant-number-BPS}%
 N_i=\sqrt2 N\epsilon\cdot \frac{\hat q_i}{L}=N^{1/3}\ \frac{\hat
 q_i}{L}\ ,
\ee%
and therefore, %
\be\label{Delta-J-N2N3N4}%
 \Delta-\sum_i J_i=\frac{4N_2N_3N_4}{3}
\cdot \frac{\hat \mu}{\hat \mu_c}.%
\ee%

Finally, let us consider the rotating case of section
\ref{near-BPS-rot-section}, where besides $J_2,\ J_3$ and $J_4$ we
have also turned on the fourth $R$-charge $J_1$,
\be%
J_1=\frac{4\sqrt2}{3}\ N^{3/2}\epsilon^3\ \cdot\frac{1}{L}
\sqrt{\hat q_1(\hat q_1+\hat\mu)} =\frac{2}{3}\ N\cdot \frac{1}{L}
\sqrt{\hat q_1(\hat q_1+\hat\mu)}\ .
\ee%
As we see $\Delta-\sum_{i=2,3,4} J_i\sim J_1\sim
N^{3/2}\epsilon^3\sim N\to \infty$.  Instead of
$\Delta-\sum_{i=2,3,4} J_i$ it is
more appropriate to define another positive definite quantity:%
\be\label{Delta-J-4charge-BPS}%
 \Delta-\sum_{i=1}^4
J_i=N \cdot \left(\frac{\hat\mu+2\hat q_1- \sqrt{(\hat\mu
+2\hat q_1)^2-\hat\mu^2}}{3L}\right)\ \geq 0\ . %
\ee%
It is remarkable that the above BPS bound is exactly the bound on
the rotating BTZ parameters, $M_{BTZ}- J_{BTZ} \geq -1$,  in which
it becomes a sensible geometry in string theory \cite{FGMS-1}. This
bound is more general than just the extremality bound of the
rotating BTZ black hole $M_{BTZ}\geq J_{BTZ} \geq 0$. This bound
besides the rotating black hole cases also includes the case in
which we have a conical singularity which could be resolved in
string theory (\emph{cf.} Appendix B and section 5 of
\cite{FGMS-1}). We will comment on this point further in section
\ref{2d-CFT-NBPS-section}.

\subsubsection{The near-horizon far-from-BPS limit, ${\cal N}=8$ $3d$
CFT description}\label{NExt.-SYM-section}

Since in the near-horizon, far-from-BPS limit of
\eqref{near-ext-limit} we do not scale $\mu$ and $q_i$'s,  we expect
to deal with a sector of the $3d$ CFT in which $\Delta\sim J_i \sim
N^{3/2}$. As mentioned in \ref{near-ext-entropy-section}, $N\sim
\epsilon^{-2}$ which for convenience we choose
\be\label{Nvs.epsilon-brane-mass}%
\epsilon^2=\frac{9}{32N}.%
\ee%
To deduce the correct ``BMN-type'' combination of $\Delta$ and $J_i$
which correspond to physical observables, we  recall the way the
limit has been taken, and in particular%
\be%
 \tau={\epsilon}\ \frac{R_S}{R_{AdS_3}}\frac{t}{L},\qquad
\phi_i = \psi_i + \frac{{\tilde q}_i R_{AdS_3}}{q_i  R_S }
\frac{\tau}{\epsilon},\ i=2,3,4\ .%
 \ee%
Therefore, $-i\frac{\partial}{\partial
\psi_i}=-i\frac{\partial}{\partial \phi_i}=J_i$ and
\be\label{Delta-J-extremal}%
\begin{split}%
{\cal E}\equiv -i\frac{\partial}{\partial
\tau}=-\frac{R_{AdS_3}}{\epsilon \ R_S}\left(i
L\frac{\partial}{\partial t}+ i\sum_{i=2,3,4} \frac{\tilde
q_i}{q_i} \frac{\partial}{\partial \phi_i}\right) =
-\frac{R_{AdS_3}}{\epsilon \ R_S}\left(\Delta-\frac{3L}{4\sqrt2
N^{3/2}}\sum_{i=2,3,4} \frac{J_i^2}{q_i}\right)
\end{split}%
\ee%
The last equality  can be understood in an intuitive way. In this
case we are  dealing with massive giant gravitons which are far from
being BPS and hence are behaving like \emph{non-relativistic}
objects. They are rotating with angular momentum $J_i$ over circles
with radii $R_i$, $R_i^2=\frac{L^2}{R^2_S} q_i$ \eqref{near-ext-M6}.
Therefore, the kinetic energy of these rotating branes is
proportional to $\sum J^2_i/q_i$.

 Recalling that $\Delta$ is measuring the ``total'' energy of
the system,  ${\cal E}$ should have two parts: the rest-mass of the
system of giants and the energy corresponding to the ``internal''
excitations of the branes. We can see this explicitly from
\eqref{Delta,J} and \eqref{4d-ADM-mass},
\be\label{Delta-J-extremal-2}%
\begin{split}
{\cal E}&=\frac{4\sqrt2}{3}\
\frac{R_{AdS_3}}{R_S}\cdot{\frac{N^{3/2}}{\epsilon}}\cdot
\frac{\mu}{L}\cr &= {\cal E}_0+ \frac{R_{AdS_3}}{R_S}\cdot N
\frac{\mu_c}{L}\  M
\end{split}%
\ee%
 where have used
$\mu=\mu_c+\epsilon^2 \mu_c M$ ($M$ is related to the mass of BTZ
black hole \eqref{near-ext-BTZ}),
and%
\be\label{E0}%
{\cal E}_0=\frac{128 R_{AdS_3}R_S^2}{9 L^3}\cdot N^2.
\ee%
${\cal E}_0$  which is basically ${\cal E}$ evaluated at
$\mu=\mu_c$, is the rest-mass of the brane system.\footnote{One
should keep in mind that at the extremal point the system is not BPS
and hence the ``rest-mass'' of the  system is not simply the sum of
the masses of individual stacks of giants but also includes their
``binding energy'' (stored in the non-spherical shape of the
giants).  Nonetheless, it should still be proportional to the number
of giants times mass of a single giant. Eq.\eqref{E0}, however,
seems to suggest a simpler interpretation in terms of dual M2-brane
giants \cite{11d-superstars}. Inspired by the expression for the
$11d$ three-form flux, the system of giant M5-branes we start with,
\emph{e.g.} through SUGRA solution \eqref{11-metric-original}, may
also be interpreted  as spherical membranes wrapping  $S^2\in AdS_4$
while rotating on $S^7$. In terms of dual membrane giants, after the
limit, we are dealing with a system of M2-branes wrapping the
$S^2\in AdS_3\times S^2$ which has radius $R_S$. The mass of a
single such dual giant $m_0$ (as measured in $R_{AdS_3}$ units and
also noting the scaling of $AdS_4$ time with respect to $AdS_3$
time) is then
\[ \frac{m_0}{R_{AdS_3}/\epsilon}=T_{M2}(4\pi R_S^2)= \frac{4\sqrt2 R^2_S}{L^3}\cdot N^{1/2}.
\]
The number of dual membrane giants is  proportional to $N^{3/2}$
(this could be seen from a relation like \eqref{Ni-M5-flux}) and
hence one expects the
total ``rest-mass'' of the system $m_0$ to be proportional to $N^2 R^2_S$.}%

${\cal E}-{\cal E}_0$, which is proportional to $ M$, corresponds to
the fluctuations of the intersecting, deformed   M5-brane giants
about the extremal point. From the $11d$ viewpoint  we start with
the geometries corresponding to M5-brane giants intersecting on a
string, the string which lives in five dimensions, it is very
suggestive to associate ${\cal E}-{\cal E}_0$ to the mass of these
$5d$ strings. These strings hence correspond to five-brane-type
fluctuations of the original ``intersecting M5-brane giants''.

In sum, from the $3d$ CFT  viewpoint the sector describing the
near-horizon, far-from-BPS limit consists of operators
specified with%
 \be\label{Ext-3d-CFT-sector}
 \begin{split}
 \Delta\sim J_i\sim N^{3/2}, &\qquad  N\to \infty,\cr
 \frac{J_i}{N^{3/2}}\equiv \frac{4\sqrt2\tilde q_i}{3L}=fixed, &\qquad \frac{{\cal E}-{\cal
 E}_0}{N}=fixed\ ,
 \end{split}
 \ee%
where ${\cal E}$, ${\cal E}_0$ in  equations
\eqref{Delta-J-extremal},\eqref{Delta-J-extremal-2} and \eqref{E0}
are defined in terms of $\Delta$, $J_i$.

As discussed in section \ref{Rot-BTZ-NExt-limit-section} one may
obtain a rotating BTZ if we turn on the fourth $R$-charge in a
perturbative manner. In the $3d$ CFT language this means considering
the operators which besides being in the sector specified by
\eqref{Ext-3d-CFT-sector} carry the fourth $R$-charge $J_1$,
$J_1\sim N^{3/2}\epsilon^2\sim N^{1/2}$. Explicitly,%
\be\label{3rd-charge-Ext}%
 J_1=\frac{4\sqrt2}{3}\frac{N^{3/2}}{L^2}\epsilon^2\ \sqrt{\hat q_1\mu_c}\ .%
\ee%
One should note that in terms of the $AdS_3$ parameters,
since $\varphi=\epsilon\phi$,%
\be\label{varphi-momentum}%
{\cal J}\equiv
-i\frac{\partial}{\partial\varphi}=-i\frac{1}{\epsilon}\frac{\partial}{\partial\phi}=
 \frac{J_1}{\epsilon}=\frac{4\sqrt2}{3}\ {N^{3/2}\epsilon}\ \frac{\mu_c}{L}\
 \sqrt{\frac{\hat q_1}{\mu_c}}=N \ \frac{\mu_c}{L}\
 \sqrt{\frac{\hat q_1}{\mu_c}}\ .%
 \ee%
As we see ${\cal J}$, like ${\cal E}-{\cal E}_0$, is also scaling
like $N^2\epsilon\sim N$ in our decoupling limit. When $J_1$ is
turned on the expressions for the $\Delta$ and hence ${\cal E}$ are
modified, receiving contributions from $q_1$. These corrections,
recalling \eqref{Delta,J} and that $q_1$ scales as $\epsilon^4$
\eqref{near-ext-rot-limit}, vanish in the leading order. However,
one may still define physically interesting combinations like ${\cal
E}-{\cal E}_0\pm {\cal J}$. We will elaborate further on this point
in section \ref{2d-CFT-NExt-section}.

Before closing this subsection some comments are in order:
\begin{itemize}
\item{ A remarkable point which is seen directly from
\eqref{Delta-J-extremal} is that
 $-{\cal E}$ is negative definite, \emph{i.e.} there is an \emph{extremality bound}:
\be\label{Ext-bound}%
 \Delta- \sum_i f_i(J_i)\leq 0 .
\ee%
where
\[
f_i(J_i)= \frac{3L}{4\sqrt2 N^{3/2}} \frac{J_i^2}{q_i}. \]%
 (Note that one can
express $q_i$'s in terms of the $J_i$'s but since the explicit
expressions are not illuminating we do not present them here.)
 This could be thought of as a complement to the usual BPS bound, $\Delta-\sum_i J_i\geq
 0$.}
\item{
 We note that both ${\cal E}-{\cal E}_0$ and ${\cal J}$
scale as $N^{3/2}\epsilon\sim N$ which is the same scaling as the
black hole entropy \eqref{entropy-near-ext-4d}.}

\item{ Finally, the system of original intersecting giants is composed of
three stacks of M5-brane giants each containing $N_i=\sqrt2
N^{1/2}\frac{\tilde q_i}{L}$ branes and $N_i\sim N^{1/2}\to\infty$.}
\end{itemize}

\subsection{$2d$ CFT description}
As we showed in either of the near-BPS or far-from-BPS near-horizon
limits we obtain a spacetime which has an $AdS_3\times S^2$ factor.
This, within the AdS/CFT ideology, is suggesting that M-theory on
the corresponding geometries should have a $2d$ dual  CFT
description. In this section we elaborate on this $2d$ description.

\subsubsection{The near-BPS case}\label{2d-CFT-NBPS-section}
To find the possible dual field theory which describes our decoupled
geometries of \eqref{nearBPSmetric}, and their rotating
generalizations \eqref{near-BPS-Rot-BTZ}, we recall that the
original $11d$ background is a deformation of $AdS_4\times S^7$ by
the addition of three stacks of intersecting M5-brane giants which
intersect on a circle while also wrapping  four cycles of the $S^7$.
In the process of taking the decoupling limit, we take the volume of
these four cycles to be very large while keeping the radius of the
circle fixed. Therefore, the situation becomes  essentially the same
as the near-horizon limit of three stacks of intersecting flat
M5-branes in a flat eleven-dimensional background
\cite{Vijay-Larsen}.

A closely related system is coming from M-theory on a Calabi-Yau
($CY_3$) and three stacks of M5-branes
 wrapping holomorphic four cycles of the $CY_3$ \cite{MSW}.
The intersection of the M5-branes from the $5d$ supergravity
viewpoint is then a ($5d$ black) string. The near-horizon limit of
the geometry corresponding to the above intersecting M5-branes is
$AdS_3\times S^2\times CY_3$. M-theory on this decoupled geometry
has been conjectured to be described by the ${\cal N}=(0,4)$ 2d CFT,
the MSW CFT \cite{MSW} (see also \cite{MMT} and \cite{deBoer:2008fk}
for a more recent study and a complete list of references on the
topic).

In our case we have a very similar situation, namely an M-theory background  of the form $
AdS_3\times S^2\times {\cal C}_6$ with ${\cal C}_6$ being a (non-compact) $CY_3$, 
with the difference
that the $AdS_3$ geometry we obtain is in the global coordinates
(rather than the Poincare patch as in the MSW case
\cite{Vijay-Larsen, MSW}). Therefore it is expected that the dual CFT to M-theory on the background obtained in our near-BPS near horizon limit and that of the \cite{MSW} should be the same.
We hence conjecture that M-theory on the
$AdS_3\times S^2\times {\cal C}_6$ of section
\ref{near-BPS-rot-section} is dual to an  ${\cal N}=(0, 4)$ $2d$
CFT, the degrees of freedom of which are coming from the low energy
fluctuations of the intersecting M5-branes.\footnote{As discussed earlier the geometry that we start with corresponds to intersecting spherical M5-brane giant gravitons on the $AdS_4\times S^7$ background and relation to the case of \cite{MSW} where we have a collection of intersecting  M5-branes wrapping four-cycles on a $CY_3$ on the $11d$ flat background is not clear at first sight. We note that in our limit we focus on a region around the center of $AdS_4$ and hence do not see the $AdS_4$ asymptotics in the near horizon geometry. Moreover, we also blow up the radii of four-cycles over which the M5-brane giants are wrapping, replacing them with a flat space (which is a part of ${\cal C}_6$). Therefore, the distinction between the two cases is removed in the near horizon limit.}

Here are some further
comments regarding the conjecture:
\begin{itemize}
\item{
The geometry \eqref{nearBPSmetric} or \eqref{near-BPS-Rot-BTZ} were
obtained as the near-horizon, near-BPS limit of a $4d$ black hole or
a deformation of $AdS_4\times S^7$ and in the process of the limit
we focus on a narrow strip on $\mu_2, \mu_3, \mu_4$ directions. The
BTZ$\times S^2\times {\cal C}_6$ geometry and hence the
corresponding $2d$ CFT description is only describing the narrow
strips on the original $4d$ black hole.  Therefore, our $4d$ black
hole is described in terms of not a single $2d$ CFT, but a
collection of (infinitely many of) them. The \emph{only} property
which is different among these $2d$ CFT's is their central charge
and as is seen from the decoupled metric \eqref{near-BPS-Rot-BTZ}
they all have the same $L_0$ and $\bar L_0$.\\
The ``metric'' on the
space of these $2d$ CFT's is exactly the same as the metric on
${\cal C}_6$. As far as the entropy and the overall (total) number
of degrees of freedom are concerned, one can define an
\emph{effective central charge} of the theory which is the integral
over the central charge of the theory corresponding to each strip.
To compute the central charge we use the Brown-Henneaux central
charge formula \cite{Brown-Henneaux},
\[c=\frac{3\,R_{AdS}}{2\ G^{(3)}_N}.\]
The \emph{effective} total central charge is obtained by integrating
strip-wise $c$ over the ${\cal C}_6$. Noting that 
\[\int_{\mu_2^2+\mu_3^2+\mu_4^2\leq 1} d\mu_2^2 d\mu_3^2 d\mu_4^2=\frac16, \]
the \emph{effective} central charge of the system is
\be\label{central-charge-BPS}%
c_L=c_R=c=8N_2N_3N_4= 2N\cdot\frac{\hat\mu_c}{L}.%
\ee%
It is notable that the central charge $c\sim N\to\infty$.}

\item{The $2d$ CFT is described by $L_0, \bar L_0$
which are related to the BTZ black hole mass and angular momentum
\cite{David:2002wn}
as %
\be\label{2dvs.BTZ}%
L_0=\frac{6}{c}N_L=\frac{1}{4}({M_{BTZ}}-J_{BTZ}), \qquad
\bar L_0=\frac{6}{c}N_R=\frac{1}{4}({M_{BTZ}}+J_{BTZ}).%
 \ee%
The above expressions for $L_0, \bar L_0$ are given for
$M_{BTZ}-J_{BTZ}\geq 0$ when we have a black hole description. When
$-1\leq M_{BTZ}-J_{BTZ} <0$, we need to replace them with
$L_0=-\frac{c}{24} a^2_+$, $\bar L_0=-\frac{c}{24} a_-^2$ (in the
conventions introduced in the Appendix B of \cite{FGMS-1}) \cite{
David:2002wn, Justin-Mandal}. In the special case of \emph{global}
$AdS_3$ background, where $a_+=a_-=1/2$ formally corresponding to
$M_{BTZ}=-1, J_{BTZ}=0$,  the ground state is describing an NSNS
vacuum of the $2d$ CFT \cite{Maldacena:2000dr}.
 The expressions for $M_{BTZ}$ and $J_{BTZ}$ in terms of the
system of giants are given in \eqref{BPS-M,J} and
\eqref{cal-j-BPS}.

}
\item{With the above identification, it is readily seen that the Cardy formula for the entropy of
a $2d$ CFT%
\be\label{Cardy-entropy}%
\begin{split}%
S_{2d\ CFT} &= 2\pi\left(\sqrt{cN_L/6}+\sqrt{cN_R/6}\right)\cr%
&= \frac{\pi}{6}\
c\left(\sqrt{M_{BTZ}-J_{BTZ}}+\sqrt{M_{BTZ}+J_{BTZ}}\right)
\end{split}
\ee%
exactly reproduces the expressions for the entropy we got in the
previous section, \eqref{entropy-near-BPS-4d}, once we substitute
for the central charge from \eqref{central-charge-BPS} and
$M_{BTZ},\ J_{BTZ}$ from \eqref{BPS-M,J} and \eqref{cal-j-BPS}. 

We would like to point out that the usage of Bekenstein-Hawking entropy formulae \eqref{entropy-near-BPS-4d} and \eqref{entropy-near-ext-4d}, as well as the Cardy formula is limited to the cases with non-Planckian horizon radii and when both entropy and the central charge are both large. 
}
\item{It is also instructive to directly compare the $3d$ description discussed in
\ref{NBPS-SYM-section}  and the $2d$ field theory descriptions.
Comparing the expressions for $M_{BTZ}, J_{BTZ}$ and
$\Delta-\sum_{i=2,3,4} J_i$, $J_1$, we see that
they match; explicitly%
\be\label{2d-4d}%
\Delta-\sum_{i=2,3,4} J_i=\frac{c}{6} (M_{BTZ}+1),\qquad
J_1=\frac{c}{6} J_{BTZ}\ . %
\ee%
This is very remarkable because it makes a direct contact between
the $2d$ and $3d$ CFT descriptions. The $3d$ CFT BPS bound,
\emph{i.e.} $\Delta-\sum_{i=1,2,3,4} J_i\geq 0$ now translates into
the bound $M_{BTZ}-J_{BTZ}\geq -1$. This means that the $3d$ CFT,
besides being able to describe the rotating BTZ black holes, can
describe the conical spaces too. In other words,
$\Delta-\sum_{i=1}^4 J_i=0$ and $N\frac{\hat \mu_c}{3L}$
respectively correspond to global $AdS_3$ and massless BTZ cases and
when
\[
0<\Delta-\sum_{i=1}^4 J_i<\frac{c}{6}=N\frac{\hat \mu_c}{3L}\ ,
\]
the $3d$ CFT is describing a conical space, provided that $\gamma$,
\[
\gamma^2\equiv \frac{6}{c}\left(\Delta-\sum_{i=1}^4 J_i\right)-1,
\]
is a rational number.  One should also keep in mind that entropy and
temperature are sensible only when $\Delta-\sum_{i=1}^4 J_i\geq
\frac{c}{6}$; for smaller values the degeneracy of the operators in
the $3d$ CFT is not large enough to form a horizon of finite size
(in $3d$ Planck units).}
\end{itemize}

\subsubsection{The far-from-BPS case}\label{2d-CFT-NExt-section}
In the far-from-BPS case, the near-horizon limit results in a
background which again contained $AdS_3\times S^2$ as a subspace.
This background, however is quite different from what we obtained in
the near-BPS case, as there is a different relation between the
radii of the $AdS$ and $S$, and the $S^1\subset AdS_3$ ranges over
$[0,2\pi\epsilon]=[0,3/(4\sqrt{2N})]$. In addition, the $11d$
 metric is now a warped product of five and six dimensional
subspaces.  According to the usual AdS/CFT ideology we expect
M-theory on $AdS_3\times S^2\ltimes {\cal M}_6$ to have a dual $2d$
CFT description. Noting that this background is a non-supersymmetric
one, and also the points stressed above, this $2d$ CFT cannot be the
${\cal N}=(0, 4)$ expanded about its maximally supersymmetric ground
state. In the following we just make some general remarks on the
conjectured $2d$ CFT:
\begin{itemize}
\item{ Like the $10d$ example \cite{FGMS-1}, we expect this $2d$ CFT  to reside on the $R\times S^1$ causal boundary of
the $AdS_3\times S^2$ geometry.}
\item{One may use
the Brown-Henneaux analysis \cite{Brown-Henneaux} to compute the
central charge of this $2d$ CFT:%
\be\label{central-charge-Extremal}%
c=\frac{3R_{AdS_3}\epsilon}{2 G^{(3)}_N}= 3 \frac{\mu_c}{L\sqrt{f_0}}\ N. %
\ee%
As in the near-BPS case \eqref{central-charge-BPS} the central
charge scales as $N\to\infty$. Its expression in terms of  the
number of  five-branes in the stack is, however, more complicated
({\it cf.} \eqref{tildeq-q} and \eqref{five-brane-number}).

}
\item{The $4d$ or $3d$ black hole entropies given in
\eqref{entropy-near-ext-4d} and  \eqref{entropy-near-ext-3d} take
exactly the same form obtained from counting the number of
microstates of a $2d$ CFT, \emph{i.e.} the Cardy formula
\eqref{Cardy-entropy}, with the central charge
\eqref{central-charge-Extremal} and $M_{BTZ}$ and $J_{BTZ}$ given
in \eqref{ext-M,J}.}
\item{As discussed in section \ref{NExt.-SYM-section}, there is a
sector of ${\cal N}=8$, $d=3$ CFT which describes M-theory on the
background found in section \ref{Rot-BTZ-NExt-limit-section}. This
sector is characterized by ${\cal E}-{\cal E}_0$ and ${\cal J}$.
From \eqref{Delta-J-extremal-2} and \eqref{varphi-momentum} one can
readily express the $3d$ parameters in terms of $2d$ parameters, namely:%
\be\label{4d-2d-Extremal}%
{\cal E}-{\cal E}_0=\frac{c}{6} M_{BTZ}\ , \qquad {\cal
J}=\frac{c}{6} J_{BTZ}\ , %
\ee%
where $c$ is given in \eqref{central-charge-Extremal} and
$M_{BTZ}, J_{BTZ}$ are given in \eqref{ext-M,J}. The above
relations have the form of the near-BPS case discussed in section
\ref{2d-CFT-NBPS-section}. Note, however, that in this case ${\cal
E}-{\cal E}_0$ is measuring the mass of the BTZ with the zero
point energy set at the massless BTZ case (rather than global
$AdS_3$).}

\end{itemize}

\section{Discussion}

In this paper we studied near-horizon decoupling limits of
near-extremal \emph{three-charge} black holes of $U(1)^4$ $4d$
gauged SUGRA; a parallel analysis for the two-charge black holes of
$U(1)^3$ $5d$ gauged SUGRA was carried out in \cite{FGMS-1}, see
also \cite{Balasubramanian:2007bs}. We showed that there are two
such near-extremal limits, the near-BPS and the far-from-BPS limits.
In both cases the eleven-dimensional uplift of the $4d$ black holes
lead to space-times with $X_{M,J}\times S^2$ factors, $X_{M,J}$
being a rotating BTZ black hole. As the eleven-dimensional uplift of
the $4d$ black holes are deformations around $AdS_4\times S^7$
geometry, the M-theory on the geometries obtained after the
near-horizon limit and the process of taking the decoupling limit
should have a description in terms of the $3d$ ${\cal N}=8$  CFT. As
we argued taking the near-horizon limit corresponds to working in
BMN-type sectors of large R-charges in this $3d$ CFT. Moreover,
appearance of the $AdS_3$ (or rotating BTZ) factors indicates that
there should be a description in terms of $2d$ CFT's, the central
charge and $L_0$ and $\bar L_0$ of which we identified in terms of
the rotating BTZ parameters. In the near-BPS case this $2d$ CFT
description is closely related to the three-charge black holes
coming in the near horizon limits of three stacks of intersecting
M5-branes \cite{MSW, deBoer:2008fk}. In our case, however, the
M5-brane picture originates not from flat M5-branes wrapping
holomorphic four cycles of $CY3$ over which the M-theory is
compactified, but from spherical M5-branes, M5-brane giants,
wrapping five-spheres in $S^7$ of the original $AdS_4\times S^7$
geometry. In other words, the $2d$ CFT in the near-BPS case is a
specific sector of the $6d$ ${\cal N}=(0,2)$ CFT on $R\times
S^1\times \Sigma_4$, where $\Sigma_4$ is a four-dimensional space
over which the M5-branes overlap.

Our knowledge of M-theory and the ${\cal N}=8$ $3d$ CFT are both
very limited, so identification of sectors in the $3d$ CFT which
describes the M-theory on the above backgrounds involving
$AdS_3\times S^2$ factors is not very illuminating. Nonetheless, one
can use our results to learn more about, at least specific
\emph{decoupled} sectors, of the $3d$ CFT using the better
understood $2d$ CFT's. On the other hand, M-theory in the Discrete
Light-Cone Quantization (DLCQ) is conjectured to be described by
Matrix models \cite{BMN, DSV, BFSS}. In particular, it has been
argued that in the DLCQ description of M-theory on the $AdS_4\times
S^7$ and on the corresponding $11d$ plane-wave are the same
\cite{DLCQ-Penrose, TGMT}. It is then reasonable to look for a
Matrix theory description of the sectors of the $3d$ CFT or the $2d$
CFT's we have identified. Apart from specific sectors in the BMN (or
plane-wave) matrix model one may also search for matrix model
description of (DLCQ of) M-theory on the geometries obtained after
the near-horizon limits. In the same spirit as the BMN matrix model
\cite{DSV, TGMT} this matrix model is presumably coming from
``quantization'' of spherical M2-branes on the corresponding
background.

As we argued the $X_{M,J}\times S^2$ factors in both the near-BPS
and far-from-BPS cases are solutions to five-dimensional
supergravities; the near-BPS case is a solution to the
\emph{ungauge} $U(1)^3$ SUGRA, the STU model, while the far-from-BPS
case is a solution to the \emph{gauged} $U(1)^3$ SUGRA. As discussed
the $AdS_3\times S^2$ obtained in the near-BPS case can also be
obtained from the near-horizon limit of magnetically charged string
solutions of the STU model, with the important difference that in
our limit we obtain $AdS_3$ in global coordinates rather than the
Poincare patch of \cite{Sabra-Klemm}. In the far-from-BPS limit, it
is  an open question to check if our $AdS_3\times S^2$ can be
obtained as the near-horizon limit of a $5d$ string solution. As the
$AdS_3\times S^2$ in this case is not a supersymmetric solution,
this $5d$ string solution, if it exists, is expected to be a
non-supersymmetric solution. Searching for such a $5d$ string
solution is an interesting open question because, this solution, if
it exists, should be a circular string solution in the $AdS_5$
background. As such, one then expects to have another description in
terms of sectors of ${\cal N}=4$, $4d$ SYM.

As we showed in section 5, the $AdS_3\times S^2$ solution coming
from the far-from-BPS limit is a solution to $5d$ $U(1)^3$ gauged
SUGRA with three magnetic fluxes over the $S^2$. Recalling that this
$AdS_3\times S^2$ is a part of a solution to $11d$ SUGRA with metric
and three-form \eqref{near-ext-metric} and \eqref{near-ext-3form},
it is then plausible to expect that the $5d$ gauged $U(1)^3$ theory
could be obtained from warped reduction of $11d$ SUGRA over
\eqref{near-ext-M6}. Explicitly, we expect the metric reduction
ans\"atz to be%
\be\label{11d-reduction-metric}%
 ds^2_{(11)} = \Delta^{\frac43} g^{(5)}_{\mu\nu}(x) dx^\mu dx^\nu +
 \Delta^{-\frac23} \sum_{i=2,3,4} X^{-1}_i\
 \left(d\mu_i^2+\mu_i^2(d\psi_i+L\ A_i)^2\right)
\ee%
where $x_\mu$ denote the five-dimensional coordinates and $A_i$ are
the three $U(1)$ gauge fields. $X_i$, which are constrained by
$X_2X_3X_4=1$,  constitute the two scalars of the $5d$ $U(1)^3$
gauged SUGRA. For the case with vanishing electric charges, like our
$AdS_3\times S^2$ we expect
$\Delta=\mu_1=\sqrt{1-(\mu_2^2+\mu_3^2+\mu_4^2)}$. In \cite{FGMS-1}
we proposed a similar reduction of IIB SUGRA to a six-dimensional
$U(1)^2$ gauged SUGRA. Verifying the consistency of this reduction
and completing the reduction ans\"atz for the three-from is
postponed to feature works \cite{progress-6d}. Although for the
cases where $\Delta=\mu_1$, $\mu_1=0$ is a curvature singularity we
expect this singularity to be resolved once the quantum gravity
(M-theory) effects are taken into account. If the above proposal for
obtaining the $U(1)^3$ $5d$ gauged SUGRA is verified, it will be
very interesting to see if there is a direct relation between the
other reduction which leads to the same $5d$ theory, \emph{i.e.}
reduction of $10d$ IIB theory on $S^5$
\cite{Cvetic:1999xp,Duff:1999rk}.

\section*{Acknowledgments}

We would like to thank Wafic Sabra for useful correspondence.

\appendix

\section{ Near-Horizon Geometry as a Solution to $11d$ SUGRA}\label{appendixA}

Here, we present the details which establish that the  near-horizon
far-from-BPS limit of the three-charge black hole obtained in
section $3.2$ and the perturbatively added four-charge black hole
obtained in section $3.3.2$ are solutions to the eleven-dimensional
supergravity equations of motion:
\begin{eqnarray}\label{C1}
R_{M\,N}-\frac12 \,R~g_{M\,N} &=&  T_{M\,N}, \\ \label{C2} d \ast F
+\frac12 F \wedge F &=& 0,
\end{eqnarray}
where $T_{M\,N}$ is the energy-momentum tensor of the flux
\begin{equation}\label{C21}
T_{M\,N} = \frac{1}{12}\left( F_{M\,P\,Q\,R}\,F_N^{~~P\,Q\,R} -
\frac18 F_{P\,Q\,R\,S}F^{P\,Q\,R\,S} \right).
\end{equation}
The metric for the three-charge case is given in equations
\eqref{near-ext-metric}, \eqref{near-ext-BTZ} and
\eqref{near-ext-M6} and that for the four-charge case is given in
\eqref{near-ext-metric}, \eqref{near-BPS-Rot-BTZ} and
\eqref{near-ext-M6}. The three-form flux for both the three-charge
case and the four-charge case is the same, given in equation
\eqref{near-ext-3form}; its  four-form field strength is given by,
\begin{equation}\label{C3}
\begin{split}
F^{(4)} = - L^2 q_2\left(1+\frac{4\,q_3\,q_4}{L^2} \right)^{\frac12}
\mu_2\,d\mu_2\wedge d\psi_2 \wedge d^3\Omega_2 - L^2
q_3\left(1+\frac{4\,q_4\,q_2}{L^2} \right)^{\frac12}
\mu_3\,d\mu_3\wedge d\psi_3 \wedge d^3\Omega_2\\- L^2
q_4\left(1+\frac{4\,q_2\,q_3}{L^2} \right)^{\frac12}
\mu_4\,d\mu_4\wedge d\psi_4 \wedge d^3\Omega_2
\end{split}
\end{equation}
and it's Hodge dual,
\begin{equation}\label{C4}
\begin{split}
\star ~F^{(4)} = - \frac{L^4}{q_2} \left(1+\frac{4\,q_3\,q_4}{L^2}
\right)^{\frac12} { \rho} \, \mu_3\,\mu_4\, d{ \rho} \wedge d{ \tau}
\wedge d\varphi \wedge d\mu_3 \wedge d\psi_3 \wedge d\mu_4 \wedge
d\psi_4 \\ - \frac{L^4}{q_3} \left(1+\frac{4\,q_4\,q_2}{L^2}
\right)^{\frac12} { \rho} \, \mu_4\,\mu_2\, d{ \rho} \wedge d{ \tau}
\wedge d\varphi \wedge d\mu_4 \wedge d\psi_4\wedge d\mu_2 \wedge
d\psi_2 \\ - \frac{L^4}{q_4} \left(1+\frac{4\,q_2\,q_3}{L^2}
\right)^{\frac12} { \rho} \, \mu_2\,\mu_3\, d{ \rho} \wedge d{ \tau}
\wedge d\varphi \wedge d\mu_2 \wedge d\psi_2 \wedge d\mu_3 \wedge
d\psi_3.
\end{split}
\end{equation}
It is clear from \eqref{C4} and \eqref{C3} that the two terms in
\eqref{C2} separately vanish and thus the flux equation of motion is satisfied. The non-vanishing components of the energy-momentum tensor \eqref{C21} for the three-charge case are:
\begin{equation}\label{C5}
\begin{split}
\frac{T_{{ \tau}\,{ \tau}}}{g_{{ \tau}\,{ \tau}}} = \frac{T_{{
\rho}\,{ \rho}}}{g_{{ \rho}\,{ \rho}}}
=\frac{T_{\varphi\,\varphi}}{g_{\varphi\,\varphi}} = - \frac{f_0 +
2}{4\,(q_2\,q_3\,q_4)^{\frac23}\,\mu_1^{\,\frac43}},\\
\frac{T_{i\,j}}{g_{i\,j}} = \frac{f_0 +
2}{4\,(q_2\,q_3\,q_4)^{\frac23}\,\mu_1^{\,\frac43}}, \qquad
\frac{T_{\mu_2\,\mu_2}}{g_{\mu_2\,\mu_2}} =
\frac{T_{\psi_2\,\psi_2}}{g_{\psi_2\,\psi_2}} = -\frac{f_0
-\frac{8\,q_3\,q_4}{L^2}}{4\,(q_2\,q_3\,q_4)^{\frac23}\,\mu_1^{\,\frac43}},\\
\frac{T_{\mu_3\,\mu_3}}{g_{\mu_3\,\mu_3}} =
\frac{T_{\psi_3\,\psi_3}}{g_{\psi_3\,\psi_3}} = -\frac{f_0
-\frac{8\,q_4\,q_2}{L^2}}{4\,(q_2\,q_3\,q_4)^{\frac23}\,\mu_1^{\,\frac43}},
\qquad  \frac{T_{\mu_4\,\mu_4}}{g_{\mu_4\,\mu_4}} =
\frac{T_{\psi_4\,\psi_4}}{g_{\psi_4\,\psi_4}} = -\frac{f_0
-\frac{8\,q_2\,q_3}{L^2}}{4\,(q_2\,q_3\,q_4)^{\frac23}\,\mu_1^{\,\frac43}},
\end{split}
\end{equation}
where $g_{i\,j}$ is the metric on the two-sphere. Although the flux
for the four-charge case is the same as for the three-charge case,
since the metric has one extra component, the energy-momentum tensor
for the four-charge case has one more component, apart from
\eqref{C5}
\begin{equation}\label{C6}
\frac{T_{{ \tau}\,\varphi}}{g_{{ \tau}\, \varphi}} = - \frac{f_0 +
2}{4\,(q_2\,q_3\,q_4)^{\frac23}\,\mu_1^{\,\frac43}}.
\end{equation}
The Ricci tensor of the three-charge metric \eqref{near-ext-metric}, \eqref{near-ext-BTZ} and
\eqref{near-ext-M6} has the following non-vanishing components:
\begin{equation}\label{C7}
\begin{split}
\frac{R_{{ \tau}\,{ \tau}}}{g_{{ \tau}\,{ \tau}}} = \frac{R_{{
\rho}\,{ \rho}}}{g_{{ \rho}\,{ \rho}}}
=\frac{R_{\varphi\,\varphi}}{g_{\varphi\,\varphi}} = - \frac{f_0 +
2}{6\,(q_2\,q_3\,q_4)^{\frac23}\,\mu_1^{\,\frac43}},\\
\frac{R_{i\,j}}{g_{i\,j}} = \frac{f_0 +
2}{3\,(q_2\,q_3\,q_4)^{\frac23}\,\mu_1^{\,\frac43}}, \qquad
\frac{R_{\mu_2\,\mu_2}}{g_{\mu_2\,\mu_2}} =
\frac{R_{\psi_2\,\psi_2}}{g_{\psi_2\,\psi_2}} = -\frac{f_0 -1
-\frac{12\,q_3\,q_4}{L^2}}{6\,(q_2\,q_3\,q_4)^{\frac23}\,\mu_1^{\,\frac43}},\\
\frac{R_{\mu_3\,\mu_3}}{g_{\mu_3\,\mu_3}} = -\frac{f_0 -1
-\frac{12\,q_4\,q_2}{L^2}}{6\,(q_2\,q_3\,q_4)^{\frac23}\,\mu_1^{\,\frac43}},\qquad
\frac{R_{\mu_4\,\mu_4}}{g_{\mu_4\,\mu_4}} =
\frac{R_{\psi_4\,\psi_4}}{g_{\psi_4\,\psi_4}} = -\frac{f_0 -1
-\frac{12\,q_2\,q_3}{L^2}}{6\,(q_2\,q_3\,q_4)^{\frac23}\,\mu_1^{\,\frac43}}.
\end{split}
\end{equation}
 The Ricci tensor for the four-charge metric \eqref{near-ext-metric}, \eqref{near-BPS-Rot-BTZ} and
 \eqref{near-ext-M6} is \eqref{C7} plus one more component:
 \begin{equation}\label{C8}
 \frac{R_{{ \tau}\,\varphi}}{g_{{ \tau}\, \varphi}} = - \frac{f_0 + 2}{6\,(q_2\,q_3\,q_4)^{\frac23}\,\mu_1^{\,\frac43}}.
 \end{equation}
Using the Ricci scalar
\begin{equation}\label{C9}
R =  \frac{f_0 + 2}{6\,(q_2\,q_3\,q_4)^{\frac23}\,\mu_1^{\,\frac43}}
\end{equation}
and plugging  \eqref{C7} and \eqref{C5} into \eqref{C1}, it is clear
that the near-horizon far-from-BPS limit of the three-charge black
hole is indeed a solution to eleven-dimensional supergravity.
Similarly plugging \eqref{C7}, \eqref{C8} and \eqref{C5}, \eqref{C6}
into \eqref{C1} proves that the small-charge near-horizon
far-from-BPS limit of the four-charge black hole is a solution the
eleven-dimensional supergravity.

\section{The Entropy Function Analysis}\label{appendixB}

To study thermodynamic property of both of the near-BPS and
 far-from-BPS black holes one may use Sen's
entropy function \cite{Sen-review}, see also
\cite{Morales:2006gm,Dimitru-Yavar}. The procedure and computations
are essentially the same as the one presented in \cite{FGMS-1} for
the BTZ$\times S^3$ geometries. To run the entropy function
machinery (for a review see \cite{Sen-review}) we plug the ans\"atz
\eqref{ansatz-bps-fields} into the entropy function $F$:
\begin{equation}\label{F-def}
\begin{split}
    F(R_A, R_S, u^i; p^i)=\frac{1}{16 G^{(5)}_N}\int\,dx^H\,\sqrt{-g^{(5)}}
    \Big(F_{\tau\mu}
    \frac{\partial \mathcal{L}}{\partial F_{\tau\mu}}-
    \mathcal{L}\Big)=\frac{-1}{16
    G^{(5)}_N}\int\,dx^H\,\sqrt{-g^{(5)}} \ \mathcal{L}\ .
\end{split}
\end{equation}
In writing the second equality we have used the fact that for our
ans\"atz \eqref{ansatz-bps-fields} we do not have electric
charges/flux ($F_{\tau\mu}=0$).  We take ${\cal L}$ to be the
ungauged $5d$ supergravity Lagrangian \eqref{STU-ungauged-action}
for the near-BPS case and the gauged  $5d$ supergravity Lagrangian
\eqref{gauged-action} for the far-from-BPS case. $\{x^H\}$ denotes
the three-dimensional horizon of the $5d$ black string solution
which for our case it is $S^1\times S^2$, where $S^1$ is a circle of
radius $\rho_h$.

According to the entropy function procedure \cite{Sen-review}, the
minimum value of the above entropy function is equal to the entropy
of the corresponding $5d$ near-extremal black hole. The entropy
function is minimized on the solutions of the ``field'' equations
\begin{subequations}\label{eom-entropy function}
\begin{align}
\frac{\partial F(R_A,R_S,u^i;p^i)}{\partial R_A}=0\ &,\qquad
\frac{\partial F(R_A,R_S,u^i;p^i)}{\partial R_S}=0\\
\frac{\partial f(R_A,R_S,u^i;p^i)}{\partial u^i}&=0, \quad i=1,2.
\end{align}
\end{subequations}
Note that $u^3=(u^1u^2)^{-1}$.

\subsection{The near-BPS case}

Evaluating the entropy function for the near-BPS case using the
ans\"atz \eqref{ansatz-bps-fields} we
obtain%
\begin{equation}\label{simplify-entropy function-bps}\begin{split}
     F(R_A,R_S,u^i;p^i) =\frac{-1}{16\pi G_N^{(5)}}\int
     \sin\theta d\theta d\phi\ d\phi_1\ R_A^3 R_S^2\rho_h\ \biggl[\frac{2}{R^2_S}-\frac{6}{R^2_A}
 -\frac{1}{2R_S^4}\left(\bigl(\frac{p^1}{u^1}\bigr)^2+\bigl(\frac{p^2}{u^2}\bigr)^2+
     \bigl(\frac{p^3}{u^3}\bigr)^2\right)\biggr]
\end{split}\end{equation}%
 where $u^1u^2u^3=1$  and $\phi_1,\ \phi,\
\theta$ parameterize the $S^1\times S^2$ horizon. The field
equations  \eqref{eom-entropy function}  take the form
 \begin{subequations}\label{eom-bps}
 \begin{align}
 \frac{1}{R^2_S}-\frac{1}{R^2_A}-\frac{1}{4R_S^4}&\left(\bigl(\frac{p^1}{u_1}\bigr)^2+
 \bigl(\frac{p^2}{u^2}\bigr)^2+(u^1u^2p^3)^2\right)=0\\
 -\frac{6}{R_A^2}+\frac{1}{2R_S^4}&\left(\bigl(\frac{p^1}{u^1}\bigr)^2+
 \bigl(\frac{p^2}{u^2}\bigr)^2+(u^1u^2p^3\bigr)^2\right)=0\\
 \frac{p^1}{u^1}&=u^1u^2p^3\\
 \frac{p^2}{u^2}&=u^1u^2p^3\ .
 \end{align}
 \end{subequations}
Solutions to these equations are given by \eqref{solution-bps},
\eqref{Radius-pi} and \eqref{pi-qi}.

The minimum value of the entropy function for given charges $q^i$ and $\rho_h$ is then,%
\be\label{ent-func-min}%
 F_{min}^{Near-BPS}=\frac{4\pi}{G^{(5)}_N}\ R^3_S\rho_h= {32\sqrt2}\ (N\epsilon)^{3/2}\ \frac{\hat\mu_c}{L^3}\cdot \rho_h
 \ \mu_2^0\mu_3^0\mu_4^0 %
\ee%
where in the second equality we have used \eqref{STU-GN} and
\eqref{pi-qi}. Upon integration over the angles $\mu_2,\ \mu_3$ and
$\mu_4$, we obtain the ``total'' entropy of the system which has
exactly the same value as the $3d$ rotating BTZ black hole
\eqref{entropy-near-BPS-3d}, or the original $4d$ black hole
\eqref{entropy-near-BPS-4d}.

\subsection{The far-from-BPS case}

The entropy function for the far-from-BPS case (when we have $5d$
gauged SUGRA) is%
\begin{equation}\label{simplify-entropy function-extremal}
\begin{split}
     F(R_A,R_S,u^i;p^i)=-\frac{1}{16\pi G_N^{(5)}}\int
     \sin\theta d\theta d\phi d\phi_1\, R_A^3 R^2_S \rho_h\ \Bigg[&\frac{2}{R_S^2}-\frac{6}{R_A^2}
     +\frac{4}{L^2}\left(\frac{1}{u^1}+\frac{1}{u^2}+\frac{1}{u^3}\right)\\&-\frac{1}{2R_S^4}
     \biggl(\bigl(\frac{p^1}{u^1}\bigr)^2+\bigl(\frac{p^2}{u^2}\bigr)^2+
     \bigl(\frac{p^3}{u^3}\bigr)^2\biggr)\Bigg]
\end{split}
\end{equation}
The field equations after some simplification take the form
\begin{subequations}\label{eom-exremal}
\begin{align}
-\frac{2}{R_A^2}+\frac{2}{R_S^2}+\frac{4}{L^2}\left(\frac{1}{u^1}+\frac{1}{u^2}+\frac{1}{u^3}
\right)-\frac{1}{2R_S^4}\left(\bigl(\frac{p^1}{u^1}\bigr)^2+\bigl(\frac{p^2}{u^2}\bigr)^2+
     \bigl(\frac{p^3}{u^3}\bigr)^2\right)&=0\\
-\frac{4}{R_A^2}-\frac{2}{R_S^2}+\frac{1}{R_S^4}
\left(\bigl(\frac{p^1}{u^1}\bigr)^2+\bigl(\frac{p^2}{u^2}\bigr)^2+
     \bigl(\frac{p^3}{u^3}\bigr)\right)&=0\\
\frac{4}{L^2}\left(-\frac{1}{u^1}+u^1u^2\right)-\frac{1}{R_S^4}\left(-\bigl(\frac{p^1}{u^1}\bigr)^2
+(u^1u^2p^3)^2\right)&=0\\
\frac{4}{L^2}\left(-\frac{1}{u^2}+u^1u^2\right)-\frac{1}{R_S^4}\left(-\bigl(\frac{p^2}{u^2}\bigr)^2
+(u^1u^2p^3)^2\right)&=0 \ .
\end{align}
\end{subequations}
The solution to the above set of equations is given in
\eqref{exremal-as-5d-solution} and \eqref{def-Q-i}.

The minimum value of the entropy function for a given set of $p^i$ and $\rho_h$, after substituting for $G^{(5)}_N$
from \eqref{gaugedU(1)3-GN}, is %
\be%
F_{min}^{\textrm{far-from-BPS}}=\frac{8\sqrt2}{3}\cdot N^{3/2}\epsilon\  \frac{R^2_SR_A}{L^3}\ \rho_h %
\ee%
which is the same expression as \eqref{entropy-near-ext-4d} once we
replace for the $R_S$ and $R_A$ and $\rho_h$ in terms of the charges
and $M, J$.

\end{document}